%% file: main.tex
\documentclass[10pt, conference]{IEEEtran}

\usepackage{booktabs} 
\usepackage{wrapfig}
\usepackage{todonotes}
\usepackage{amsmath}
\usepackage{amssymb}
\usepackage{algorithm}
\usepackage{algpseudocode}
\usepackage{subcaption}
\usepackage{fancyhdr}
\usepackage{tikz}
\usepackage[skins]{tcolorbox}
\usetikzlibrary{matrix,positioning,mindmap,automata,fit,backgrounds,
shapes, arrows,petri,calc, decorations.pathreplacing}

\makeatletter
\def\BState{\State\hskip-\ALG@thistlm}
\makeatother

\algblockdefx[Machine]{BeginMachine}{EndMachine}%
[1]{\textbf{machine} #1 \{}%
{\}}
\algblockdefx[StartState]{BeginStartState}{EndStartState}%
[1]{\textbf{start state} #1 \{}%
{\}}
\algblockdefx[State]{BeginState}{EndState}%
[1]{\textbf{state} #1 \{}%
{\}}
\algblockdefx[On]{BeginOn}{EndOn}%
[1]{\textbf{on} #1 \textbf{do} \{}%
{\}}
\algblockdefx[Entry]{BeginEntryn}{EndEntry}%
[1]{\textbf{entry} \{}%
{\}}
\algblockdefx[For]{BeginFor}{EndFor}%
[3]{\textbf{for} #1 $\gets$ #2 \textbf{to} #3 \textbf{do}}%
{\textbf{end for}}
\algblockdefx[Foreach]{BeginForeach}{EndForeach}%
[1]{\textbf{for each} #1 \textbf{do}}%
{\textbf{end for}}
\algblockdefx[Fun]{BeginFun}{EndFun}%
[2]{\textbf{fun} #1(#2) \{}%
{\}}

\newcommand{\vect}[1]{\boldsymbol{#1}}

\newcommand{\naturals}{\mathbb{N}}
\newcommand{\policy}{\mathcal{P}}

\newcommand{\NodeDivision}{\mathsf{NodeDivision}}
\newcommand{\PathPlanning}{\mathsf{PathPlanning}}
\newcommand{\AStar}{\mathsf{A^*}}
\newcommand{\PID}{\mathsf{PID}}

\newcommand{\goal}{\mathit{goal}}
\newcommand{\Risk}{\mathsf{Risk}}
\newcommand{\ReplanningInterval}{\mathsf{ReplanningInterval}}
\newcommand{\ID}{\mathsf{id}}

\newcommand{\ND}{\mathsf{ND}}
\newcommand{\OP}{\mathsf{OP}}

\begin{document}

%
%

\pagestyle{plain}
\pagenumbering{arabic}

\fancyfoot{Hello}

\input{abs}
\input{intro}
\input{problem}

\input{model}
\input{overall}
\input{algorithm}

\input{experiments}
\input{related}
\input{concl}

\bibliographystyle{IEEEtran}
\bibliography{bibliography}
\end{document}

%% file: abs.tex
\title{Decentralized Multi-UAV Routing in the Presence of Disturbances}

\author{
\IEEEauthorblockN{Wei Zhao}
\IEEEauthorblockA{Department of Computing and Software\\
McMaster University, Canada\\
Email: \sf{zhaow9@mcmaster.ca}}
\and
\IEEEauthorblockN{Borzoo Bonakdarpour}
\IEEEauthorblockA{Department of Computing and Software\\
McMaster University, Canada\\
Email: \sf{borzoo@mcmaster.ca}}
}

\date{}
\maketitle

\begin{abstract}

We introduce a decentralized and online path planning technique
for a network of unmanned aerial vehicles (UAVs) in the presence of weather 
disturbances. In our problem setting, the group of UAVs are required to 
collaboratively visit a set of goals scattered in a 2-dimensional area. Each UAV 
will have to spend energy to reach these goals, but due to unforeseen 
disturbances, the required energy may vary over time and does not necessarily 
conform with the initial forecast and/or pre-computed optimal paths. Thus, we 
are dealing with two fundamental interrelated problems to find a global optimum 
at each point of time: (1) energy consumption prediction based on disturbances 
and, hence, online path replanning, and (2) distributed agreement among all UAVs 
to divide the remaining unvisited goals based on their positions and energy 
requirements. Our approach consists of four main components: (i) a distributed 
algorithm that periodically divides the unvisited goals among all the UAVs based 
on the current energy requirements of the UAVs, (ii) a local (i.e., UAV-level) 
$\AStar$-based algorithm that computes the {\em desirable} path for each UAV to 
reach the nodes assigned to it, (iii) a local PID controller that {\em predicts} 
the inputs to the UAV (i.e., thrust and moments), and (iv) a planner 
that computes the required energy and the replanning time period. We validate 
our proposed solution through a rich set of simulations and show that our 
approach is significantly more efficient than a best-effort algorithm that 
directs each idle UAV to visit the closest unvisited goal. 

\end{abstract}

%% file: intro.tex
\section{Introduction}

{\em Unmanned aerial vehicles (UAVs)}, commonly known as {\em drones}, are 
currently being used in many domains and are expected to play a significant 
role in future technologies. UAVs have application in emergency-response, 
search and rescue, transportation, border patrolling, agriculture, 
topographical surveys and inspection, and even in the entertainment industry. 
Moreover, in today's highly connected world, it is very realistic to employ 
UAVs as mobile sensors and actuators, becoming an essential part of the 
Internet of Things (IoT) and playing a crucial role in our daily lives. Each 
UAV application has its specific challenges, but they all share one severe 
restriction, namely, energy constraints that limit their use to relatively 
small-scale applications. For instance, the economically feasible coverage area 
for a topographical survey application is now limited to areas less than $1 
km^2$. This and similar scenarios call for transition from single UAV/single 
operator approaches to autonomous multi-UAV/single operator solutions. This is 
a highly challenging task as coordination of multiple interacting UAVs involves 
many subtleties, from solving computationally intractable problems, e.g., the 
vehicle routing problem (VRP)~\cite{dantzig1959truck} to dealing with 
fault-tolerance and physical disturbances. 

Multi-UAV path planning has been studied from different perspectives. Solutions 
to the distributed VRP~\cite{kmb17,pfb11} are typically concerned with finding 
optimal routes in the presence of a static or dynamic set of clients and 
demands. These efforts essentially abstract away the impact of physical 
disturbances on the cost of vehicle routes. However, in reality, 
physical disturbances such as weather conditions can have profound impact on the 
local and global energy efficiency of a network of vehicles. Furthermore, to 
our knowledge, approaches that deal with environment stimuli 
(e.g.,~\cite{bmnlkp16}) are limited to control-theoretic techniques for
single-UAV applications and it is unclear how they can be extended to a 
multi-UAV setting. 

\begin{figure}[b]
 \centering
 \includegraphics[scale=1.2]{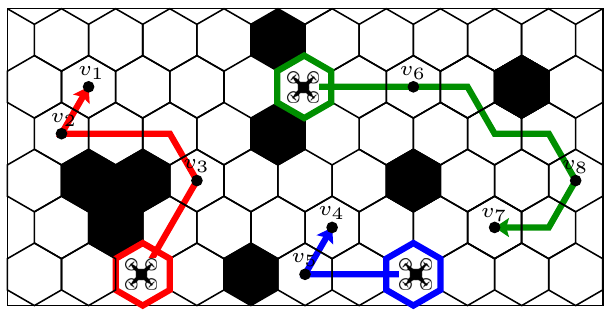}
 \caption{Hexagon grid of flight area and flight paths of UAVs.}
 \label{fig:hex}
\end{figure}

With this motivation, in this paper, we introduce a disturbance-aware 
decentralized and online path planning technique for a network of 
autonomous UAVs. Our problem setting consists of a network of quadrotor UAVs 
with peer-to-peer communication capability that are required to collaboratively 
{\em travel} to and {\em visit} a designated set of goals scattered in a 
2-dimensional area. All UAVs have a consistent map of the area that includes a 
fixed set of goals and obstacles (see Fig.~\ref{fig:hex}). That is, goals and 
obstacles do not move and new obstacles or goals do not appear or disappear from 
the area. Each UAV will have to spend energy to travel to and visit these 
goals. 
Due to unforeseen disturbances, the required energy may vary over time and does 
not necessarily conform with the initial forecast and/or pre-computed paths. 
Thus, our goal is to design an algorithm that minimizes the global energy 
consumption among all UAVs to travel to and visit all designated goals. In 
order to 
solve this optimization problem, one has to deal with two fundamental and 
interrelated subproblems to find the global optimum at each point of time: (1) 
energy consumption prediction and online path planning based on disturbances, 
and (2) distributed agreement among all UAVs to divide the remaining goals 
based on their positions.

\begin{figure}[t]
 \centering
 \includegraphics[width=\columnwidth]{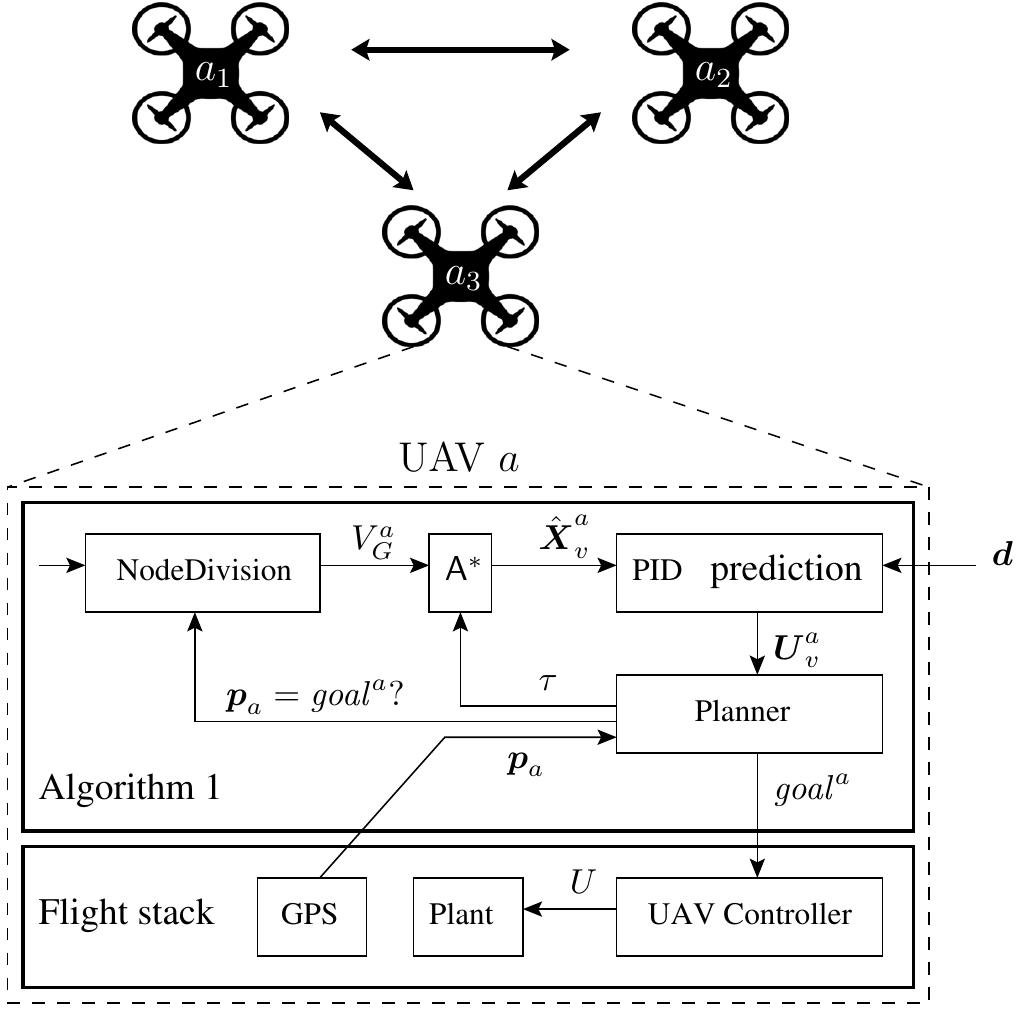}
 \caption{Building blocks of disturbance-aware decentralized path planning.}
 \label{fig:arch}
\end{figure}

Our proposed solution works as follows. Each UAV $a$ in the network runs the 
same local algorithm, which consists of four main components (see 
Fig.~\ref{fig:arch}):

\begin{itemize}
\item A synchronous distributed algorithm, called $\NodeDivision$, periodically 
divides the remaining set of goals among all the UAVs based on their current 
positions and energy requirements. This algorithm results in the set $V_G^a$ of 
nodes that the UAV sets as its goals. 

\item For each node $v \in V_G^a$, a local (i.e., UAV-level) $\AStar$-based 
algorithm~\cite{4082128} computes the {\em desirable} path $\hat{\vect{X}_v^a}$ 
from the current location of the UAV to $v$ by dividing the map into a set of 
adjacent hexagons (see Fig.~\ref{fig:hex}). This algorithm assumes that the 
current wind disturbance will continue for the foreseeable future while 
predicting energy costs. 

\item A local PID controller periodically takes the desirable path to each node 
$v \in V_G^a$ as well as the current wind disturbance vector $\vect{d}$ as 
input and estimates the {\em predicted inputs} $\vect{U}_v^a$ (i.e., 
thrust and moments) to the UAV. From this input, one can predict the 
required energy and next position of the UAV.

\item Finally, a local planner determines (1) the next goal to visit (i.e., 
$\goal^a$), which is given to the UAV flight controller, (2) the next 
replanning period $\tau$, and (3) whether the current location of the UAV 
obtained by GPS is at the intended goal.  
\end{itemize}

We validate our proposed solution through a rigorous set of simulations. First, 
we show that our technique saves $25\%$ energy on average, as compared to a 
greedy algorithm, where each idle UAV chooses to visit the closest 
unvisited goal on the flight map. We believe that given the current battery 
technology of UAVs, this is a significant improvement. Then, we analyze the 
effectiveness of different components of our algorithm and their roles in 
achieving energy savings in different scenarios. We show that in scenarios with 
a small number of goals and UAVs, the PID controller plays a crucial role in 
choosing goals and dealing with disturbances. On the contrary, for scenarios 
with larger areas, number of goals, and UAVs, periodic invocation of 
$\NodeDivision$ will result in larger energy savings. Our results are consistent 
among simulations with different number of goals and UAVs as well as different 
wind patterns. For all of our simulations, we demonstrate statistical 
significance by ensuring at least 95\% confidence interval. 

\paragraph*{Organization} The rest of the paper is organized as follows. In 
Section~\ref{sec:problem}, we formally state the problem. 
Section~\ref{sec:model} presents system model dynamics of UAVs.
Section~\ref{sec:overall} presents the overall idea of our technique, while 
Section~\ref{sec:detail} elaborates on the details. Simulation results are 
explained in Section~\ref{sec:exp}. Related work is discussed in 
Section~\ref{sec:related}. Finally, we make concluding remarks and discuss 
future work in Section~\ref{sec:concl}.

%% file: problem.tex
\section{Problem Statement}
\label{sec:problem}

We first introduce some notation. Throughout the paper, we use bold lower and 
upper case italic letters to denote vectors and matrices, respectively (e.g., 
$\vect{x}$ and $\vect{X}$). We consider discrete time, where \(t_{k}\) means 
the time at index \(k\) (\(k \geq 0\)) and \(t_{k + 1}\) is its successor with 
sampling time \({t}_{s}\), i.e., \(t_s={t}_{k+1}-t_k\). We represent a 
set of UAVs by \(A\). Let \(V\) be a set of nodes and \(V_{0}\subseteq V\) be 
the {\em depot} set, such that initially there is a one-to-one mapping from $A$ 
to $V_0$, i.e., each depot hosts at most one UAV and each UAV is initially 
assigned to a depot. Thus, for a depot $v \in V_0$, by \(v_{0}^{a}\), we mean 
the depot assigned to UAV \(a\).

We assume that each node \(v \in V\) has a 2-dimensional coordinate 
\(\vect{p}_{v} = \left\lbrack x_{v}~~y_{v} \right\rbrack^{T}\) on the 
$XY$-plane. The current position of a UAV \(a \in A\) is 
represented by
\begin{equation}
\vect{p}_{a} = \begin{bmatrix}
x_{a} & y_{a} & z_{a}
\end{bmatrix}^{T}.\label{UAVPositionDefinition}
\end{equation}
We note that throughout the paper, we will abstract away the third dimension 
$z_a$. That is, our path planning problem is only on the $XY$-plane. We 
introduce the {\em wind disturbance} \(\vect{d}\left( k \right)\) at the time 
\(t_{k}\) on the $XY$-plane, such that
\begin{equation}
\vect{d}\left( k \right) = \begin{bmatrix}
d_{x} & d_{y}
\end{bmatrix}^{T},\label{WindDisturbanceDefinition}
\end{equation}
where the \(d_{x}\) and \(d_{y}\) 
is the wind speed in \(x\)- and \(y\)-directions, respectively.

Informally, our goal is to design an online decentralized and energy-efficient 
algorithm, where a set $A$ of UAVs travel to the set $V$ of nodes in the 
presence of wind disturbances and complete a task at each node. In this paper, 
we take {\em traveling energy cost} into consideration. That is before visiting 
a node \(v\), a UAV \(a\) must travel from \(\vect{p}_{a}\) to \(\vect{p}_{v}\). 
This movement requires energy to defy gravity and time-variant wind disturbance 
\(\vect{d}\).

We now state our optimization problem formally. Given a set \(A\) of UAVs, a 
set \(V\) of nodes on the $XY$-plane and the time-variant wind disturbance 
\(\vect{d}\), our goal is to find a path plan \(\policy\) such that (1) each 
UAV $a$ at every time instance $k \in T$ chooses the next node to visit, (2) 
each node is visited only once, (3) all the nodes in $V$ are visited, and (4) 
the path eventually terminates. These constraints are formally the following:
\begin{align*}
& \policy: A \times \naturals \; \rightarrow \; V \, \cup \, \left\{ 
\perp \right\}\\
& \forall k \in T:\forall a, a' \in A: \big((a \neq a') \, \Rightarrow \,
\policy( a,k) \neq \mathcal{P}(a',k)\big)\\
& \big\{\policy(a, k)~\big\vert~a \in A \wedge k \in [0, T] \big\} 
= V\\
& \forall a \in A: \exists k \in \naturals: \forall j \geq k: \policy(a, j) = 
\perp
\end{align*}
where \(\naturals\) denotes the set of natural numbers, \(\perp\) represents 
the situation that \(\policy\) does not return an unservided node, and $T$ is 
the time index that plan \(\mathcal{P}\) terminates, i.e.,
$$T = \min\big\{k \mid \forall a \in A: \policy(a, k) = \bot\big\}.$$
 
Now, let \(V_{\mathcal{P}}^{a}\) denote the set of nodes chosen 
by $\mathcal{P}$ for vehicle $a$, i.e.,
$$V_{\mathcal{P}}^{a} = \big\{\policy(a, k) \mid k \in [0, T]\big\}.$$
Our objective is to minimize the energy needed by $\policy$:
$$\min \sum_{a \in A}\sum_{v \in V_{\policy}^{a}}\sum_{k=0}^{T}{E_{k}\left( 
a,v,\vect{d}\left( k \right) \right)}$$
where $E_{k}$ is the piece-wise traveling energy cost from time $t_{k}$ to 
$t_{k + 1}$ by UAV $a$ to node $v$ under disturbance $\vect{d}(k)$.

%% file: model.tex
\section{System Model Dynamics}
\label{sec:model}

Throughout this paper, we focus on {\em quadrotor} UAVs.

\subsection{Quadrotor Model}

\paragraph{Thrust and moment} A quadrotor has four propellers with two of 
them rotating
clockwise while the other two rotating counter-clockwise. The  
rotational speed of a propeller (denoted $\omega_i$, where $i \in [1,4]$) 
generate {\em thrust} (denoted $F_i$) and magnitude of {\em moment} (denoted 
$T_i$) in quadratic proportion~\cite{mmlk10}:
\begin{equation*}
F_{i} = \kappa_{f}\omega_{i}^{2} \;\;\;\;\;\;\;\;
T_{i} = \kappa_{m}\omega_{i}^{2}
\end{equation*}
where $\kappa_{f}$ and $\kappa_m$ are the thrust and moments proportionality 
constants, respectively.

\paragraph{Input vector} Let $l$ be the length of the arms of the 
quadrotor which is a constant value. 
The net thrust and moments on the quadrotor, which we consider as the inputs to 
the system are given by:
\begin{equation}
\begin{bmatrix}
F \\
M_{x} \\
M_{y} \\
M_{z} \\
\end{bmatrix} = \begin{bmatrix}
\kappa_{f} & \kappa_{f} & \kappa_{f} & \kappa_{f} \\
0 & l\kappa_{f} & 0 & - l\kappa_{f} \\
- l\kappa_{f} & 0 & l\kappa_{f} & 0 \\
\kappa_{m} & - \kappa_{m} & \kappa_{m} & - \kappa_{m} \\
\end{bmatrix}\begin{bmatrix}
\omega_{1}^{2} \\
\omega_{2}^{2} \\
\omega_{3}^{2} \\
\omega_{4}^{2} \\
\end{bmatrix}\label{InputComputation}
\end{equation}
where $F$ is the net thrust and $M_{x}$, $M_{y}$, and $M_{z}$ are the $X$-, 
$Y$-, and $Z$-direction moments on the quadrotor, respectively. Then, for each UAV $a \in A$, we write its inputs as
\begin{equation*}
\vect{u}_{a} = \begin{bmatrix}
F & M_{x} & M_{y} & M_{z} \\
\end{bmatrix}^{T}.
\end{equation*}
Near maximum thrust is generated when all four propellers 
approximately operate in maximum rotational speed $\omega_{\max}$. Hence, 
from \eqref{InputComputation}, we obtain:  
\begin{equation}
F_{\max} \simeq 4\kappa_{f}\omega_{\max}^{2} \label{eq:Fmax}
\end{equation}

\paragraph{State of quadrotor} Let the velocity, the roll/pitch/yaw Euler 
angles and the body frame angular velocities of a UAV $a$ be defined as the 
following:
\begin{eqnarray}
\vect{v}_{a} &=& \begin{bmatrix}
v_{x} & v_{y} & v_{z}
\end{bmatrix}^{T}\label{UAVVelocityDefinition} \\
\vect{\Theta}_{a} &=& \begin{bmatrix}
\psi & \theta & \psi
\end{bmatrix}^{T}\label{UAVEularAnglesDefinition} \\
\vect{\Omega}_{a} &=& \begin{bmatrix}
p & q & r
\end{bmatrix}^{T}. \label{UAVRotationSpeedDefinition}
\end{eqnarray}
Combining \eqref{UAVPositionDefinition}, \eqref{UAVVelocityDefinition}, 
\eqref{UAVEularAnglesDefinition}, and \eqref{UAVRotationSpeedDefinition}, we 
define the {\em state} of 
a UAV $a$ (denoted $\vect{x}_a$) by the vector
\begin{equation}
\vect{x}_{a} = \begin{bmatrix}
\vect{p}_{a}^{T} & \vect{v}_{a}^{T} & \vect{\Theta}_{a}^{T} & \vect{\Omega}_{a}^{T}
\end{bmatrix}^{T}.
\end{equation}

\subsection{Quadrotor Dynamics \& Disturbance Model}

We consider the {\em wind} as the only external disturbance. We assume 
that, at any time index $k$, the wind disturbance is modeled 
as~\eqref{WindDisturbanceDefinition}. We assume that a function $f$ at time 
index $k$ takes the state $\vect{x}_a$ of a UAV, its input $\vect{u}_a$, and 
wind disturbance $\vect{d}$, and gives the new state at time index $k+1$ as 
follows:
\begin{equation}
\vect{x}_{a}(k+1) = f\big( \vect{x}_{a}(k),\vect{u}_{a}(k), \vect{d}(k) \big). 
\label{UAVDynamics}
\end{equation}
We assume constraints on the wind disturbance, namely, maximum wind disturbance 
$\vect{d}_{\max}(k)$ and minimum wind disturbance $\vect{d}_{\min}(k)$. Thus, 
the magnitude of the wind speed is in the range $\left[ 0, d_{\max} 
\right]$ for some $d_{\max}>0$. Then, the $\vect{d}_{\max}(k)$ can be defined as 
$\|\vect{d}_{\max}(k)\|=d_{\max}$ and the direction of $\vect{d}_{\max}(k)$ and 
$\vect{v}_{a}(k)$ are opposite to each 
other, where $\left\|\cdot\right\|$ is the {\em $l^2$-norm} of a 
vector, where for vector $\vect{x} = \left[x_1~ \dots~ x_n\right]^T$, we have:
$$\left\|\vect{x}\right\| = \sqrt{\sum_{k=1}^{n}x_k^2}.$$
Similarly, the $\vect{d}_{\min}(k)$ can be defined as 
$\|\vect{d}_{\min}(k)\|=d_{\max}$ and the direction of $\vect{d}_{\min}(k)$ and 
$\vect{v}_{a}(k)$ are the same with each other~\cite{hmlo02}.

\subsection{Energy Cost of Traveling}

The main source of power consumption in a UAV is the motors of propellers, 
which is equal to
$$P_{i} = T_i\omega_i = \kappa_{m}\omega_{i}^{3},$$
for $i \in [1, 4]$. Thus, the total power consumption  is given by:
\begin{equation}
\label{eq:power}
P = \sum_{i = 1}^{4}P_{i} = \kappa_{m}\sum_{i = 1}^{4}\omega_{i}^{3}.
\end{equation}
Furthermore, the UAV has to defy wind disturbances when it is traveling to a 
node. We now compute the power needed to deal with wind disturbances during 
traveling. Given the mass $m_a$ of a UAV $a$, the force needed to defy gravity 
is $F_{g} =  m_ag$. Let 
$F_{D}$ denote the {\em drag force}~\cite{falkovich11} of wind given by:
\begin{equation}
\label{eq:drag}
F_{D} = \frac{1}{2}C_{d}\rho R\left\| \vect{v}_{a} - \vect{d} \right\|^{2}
\end{equation}
where $C_{d}$ is the drag coefficient, $\rho$ is the air density, $R$ is the 
projected area faced by the wind which is assumed to be constant for a UAV. As 
can be seen, the drag force generated by wind on a UAV is proportional to their 
relative speed.

The ground velocity of a UAV can be calculated as~$\vect{v}_a = \vect{d} + 
\vect{v}_{r}$, where $\vect{v}_r$ is the velocity relative to wind. Let 
\(\theta\) be the angle between the direction of the wind and direction from 
current location \(\vect{p}_{i}\) to the destination location \(\vect{p}_{j}\), 
where \(\vect{p}_{j} \neq \vect{p}_{i}\). In order to maintain the direction of 
UAV under wind disturbance, we must have:
\begin{equation}
\left\| \vect{v}_{a} \right\| = \sqrt{\left\| \vect{v}_{r} \right\|^2 - 
(\left\| \vect{d} \right\|\sin\theta)^{2}} + \left\| \vect{d} 
\right\|\cos\theta\label{eq:v_a,v_r,d,theta}
\end{equation}
If we let the UAV fly with maximum ground speed to achieve minimum traveling 
time, it must generate the maximum thrust force, such that:
\begin{eqnarray}
\nonumber F_{\text{Tmax}} = \sqrt{F_{\max}^{2} - F_{g}^{2}} \simeq 
\sqrt{\left( 4\kappa_{f}\omega_{\max}^{2} \right)^{2} - \left( m_ag 
\right)^{2}}.
\end{eqnarray}
Once the UAV achieves the steady flight status with maximum ground speed, we 
have \(F_{\text{Tmax}} = F_{D}\). By substituting (\ref{eq:Fmax}) and 
(\ref{eq:drag}), we have
\begin{equation}
\label{eq:v_r}
\sqrt{\left( 4\kappa_{f}\omega_{\max}^{2} \right)^{2} - \left( 
m_ag \right)^{2}} \simeq \frac{1}{2}C_{d}\rho R\left\|\vect{v}_r\right\|^{2}
\end{equation}
Finally, the total energy needed by a UAV is to travel from $\vect{p}_i$ to 
$\vect{p}_j$ can be computed as follows:
\begin{eqnarray}
E_{\left( \vect{p}_{i},\vect{p}_{j} \right)} \!\!\!\!\!\! &=& \!\!\!\!\!\! 
P \cdot t_{\left( \vect{p}_{i},\vect{p}_{j} \right)} \simeq 
4\kappa_{m}\omega_{\max}^{3}\cdot t_{\left( \vect{p}_{i},\vect{p}_{j} \right)} 
\nonumber\\
\!\!\!\!\!\! && \!\!\!\!\!\! \langle \text{substituting (\ref{eq:power})} \rangle \nonumber\\
\!\!\!\!\!\! &=& \!\!\!\!\!\! 4\kappa_{m}\omega_{\max}^{3} \cdot\frac{\left\| \vect{p}_{i} - 
	\vect{p}_{j} \right\|}{\left\| \vect{v}_a \right\|} \nonumber\\
\!\!\!\!\!\! &=& \!\!\!\!\!\! \frac{4\kappa_{m}\omega_{\max}^{3} \left\| \vect{p}_{i} - 
	\vect{p}_{j} \right\|}{\sqrt{\left\| \vect{v}_{r} \right\|^2 - (\left\| \vect{d} \right\|\sin\theta)^{2}} + \left\| \vect{d} \right\|\cos\theta} \nonumber\\
\!\!\!\!\!\! && \!\!\!\!\!\! \langle \text{substituting (\ref{eq:v_a,v_r,d,theta})} \rangle \nonumber\\
\!\!\!\!\!\! &=& \!\!\!\!\!\! \frac{4\kappa_{m}\omega_{\max}^{3}\left\| \vect{p}_{i} - \vect{p}_{j} \right\|}{\sqrt{\frac{\sqrt{\left( 4\kappa_{f}\omega_{\max}^{2} \right)^{2} - \left( m_ag \right)^{2}}}{\frac{1}{2}C_{d}\rho R} - \left(\left\| \vect{d} \right\|\sin\theta\right)^{2}} + \left\| \vect{d} \right\|\cos\theta}, \nonumber \\
\!\!\!\!\!\! && \!\!\!\!\!\! \langle \text{from}~\eqref{eq:v_r} \rangle \nonumber\\
&& \text{ } \label{eq:tenergy}
\end{eqnarray}
where $P$ is the power of the UAV when it is traveling to a node. It is not 
hard to prove that $E_{\left( \vect{p}_{i},\vect{p}_{j} \right)}$ reaches 
maximum if $\|\vect{d}\|=d_{\max}$ and $\theta=\pi$ which means the direction 
of maximum speed wind is opposite to the direction the UAV will travel to. 
Thus, in Equation~\eqref{UAVDynamics}, we denote this kind of wind at time 
index $k$ with $\vect{d}_{\max}(k)$. The impact of minimum wind 
$\vect{d}_{\min}(k)$ can be explained analogously.

%% file: overall.tex
\section{The Overall Idea}
\label{sec:overall}

\input{algomain}

We assume that the $XY$-plane is given as a {\em hexagon} grid to all the UAVs. 
Figure~\ref{fig:hex} shows a flight area, where the set of nodes to be visited 
is $\{v_1, v_2, \ldots, v_8\}$, solid black hexagons represent obstacles, and 
three UAVs are in the red, green, and blue depots. The reason we choose hexagon 
grids is that the distance between the center of a hexagon to the center of any 
of its neighboring hexagons is the same and, hence, the direction of traveling 
alone does not change the energy cost. For simplicity, we assume that each node 
$v \in V$ (including the depots) is located at the center of a hexagon. The 
main components of the algorithm are as follows (see also 
Fig.~\ref{fig:arch}).\\

\noindent \textbf{Node division.} \ Our main decentralized path planning 
technique is shown in Algorithm~\ref{alg:main}. Each UAV $a \in A$ runs the 
same algorithm. The main steps of the algorithm are as follows. In 
Line~\ref{line:nd}, the UAVs create and agree on a partition of nodes through 
communicating with each other by invoking $\NodeDivision()$. This results in a 
{\em goal set} for each UAV $a \in A$ (denoted $V_G^a$) from the current set of 
unvisited nodes (denoted $V_G$). This is a {\em synchronous} step, meaning that 
all UAVs proceed to 
Line~\ref{line:despath} simultaneously and is described in detail in Section ~\ref{subsec:nd}. For example, in Fig.~\ref{fig:hex}, the goal set of 
the leftmost red UAV is $\{v_1, v_2, v_3\}$.\\

\noindent \textbf{Computing desirable path.} \ In Line \ref{line:despath}, for 
each $v \in V_{G}^{a}$, we create a path that starts at $\vect{p}_a$ (i.e., the 
current position of $a$) and ends at $\vect{p}_v$. The path goes through a 
sequence of adjacent hexagon centers (i.e., 2D positions, called {\em 
waypoints}) by using an $\AStar$-based algorithm~\cite{4082128}. This algorithm 
(described in Section~\ref{OnlinePathPlanning}) returns a sequence of {\em 
desirable} states at each time index, such that:
\begin{eqnarray}
\hat{\vect{X}}_v^{a} &=& \nonumber \hat{\vect{x}}_a\left( t_{\vect{p}_a} \right) 
\hat{\vect{x}}
_a\left( t_{\vect{p}_a} +1\right) \hat{\vect{x}}_a\left( t_{\vect{p}_a}  +2\right)\\
&&\cdots \hat{\vect{x}}_a\left( t_{\vect{p}_a} + t_{\left( \vect{p}_{a}, \vect{p}_{v} \right)}\right) \label{Xhat_v^a}
\end{eqnarray}
where $ t_{\vect{p}_a} $ is the time index at current location of $a$, and 
$t_{\left( \vect{p}_{a}, \vect{p}_{v} \right)}$ is the desirable traveling time 
following the path generated by the $\AStar$ algorithm.\\

\noindent \textbf{Predicting energy consumption.} \ We employ a PID controller 
to predict the energy cost of the desirable path computed in 
Line~\ref{line:despath}. Thus, we apply $\PID$ (Line \ref{line:PIDpath}) to 
$\hat{\vect{X}}_v^{a}$ to compute a sequence of {\em predicted} inputs 
$\vect{U}_v^a$ of the UAV which will be used to obtain the predicted traveling 
energy cost $E_{v}^{a}$ (see 
Section~\ref{OnlinePID-basedEnergyPrediction} for details). The 
predicted energy cost of traveling to node $v$ from the current location 
of $a$ is the sum of piece-wise traveling energy cost in each sampling time 
interval during the entire path:
\begin{equation}
\label{eq:pwe}
E_v^a = \sum_{k = t_{\vect{p}_a}}^{t_{\vect{p}_a}+t_{(\vect{p}_{a}, 
\vect{p}_{v})}} E_{k}\Big(a,v,\vect{d}(k) \Big).
\end{equation}
Recall from Section~\ref{sec:problem} that $E_{k}(a,v,\vect{d}(k))$ is the 
piece-wise traveling energy cost from time $t_{k}$ to $t_{k + 1}$ by UAV $a$ 
towards node $v$ under disturbance $\vect{d}(k)$, as prescribed by 
Equation~\eqref{eq:tenergy} and based on $\vect{U}_v^{a}$. As a matter of fact, 
at time index $t_{\vect{p}_a}$, the UAV cannot get the future wind disturbance. 
Thus, we let $\vect{d}(k) = \vect{d}(t_{\vect{p}_a})$, for all $k \in 
[t_{\vect{p}_a}, t_{\vect{p}_a} + t_{(\vect{p}_{a}, \vect{p}_{v})}]$. 
Likewise, $E_{v,\max}^a$ and $E_{v,\min}^a$ can be computed by replacing 
$\vect{d}(k)$ with $\vect{d}_{\max}(k)$ and $\vect{d}_{\min}(k)$, respectively.  
Let the node that has minimum predicted traveling energy cost among all the 
nodes in $V_G^a$ be the new goal $\goal^a$ to travel to and visit 
(Line~\ref{line:goal}). For example, in Fig.~\ref{fig:hex}, for the leftmost red
UAV, $\goal^a = v_3$ in the first iteration of the algorithm. Observe that the 
union of sequences of $\goal^a$ nodes for each UAV essentially establishes plan 
$\policy$ as described in Section~\ref{sec:problem}.\\

\noindent \textbf{Replanning.} \ Next, for each node $v \in V_G^a $, 
except $\goal^{a}$, from the total predicted 
traveling energy cost and the maximum/minimum traveling energy costs, we compute 
a risk value (Line~\ref{line:risk}) and select the maximum among them 
($r_{\mathit{max}}$) to compute the time interval $\tau$ before the next 
planning procedure (Line $\ref{line:tau}$) by calling $\ReplanningInterval()$ 
(described in Section \ref{subsec:rp}). At this point, the UAV has its 
goal node. If the UAV's current location is the same as the location of its goal 
node (Lines~\ref{line:beginif} -- \ref{line:endif}), then (i) $a$ has visited 
its goal, (ii) removes $goal^a$ from $V_G^a$, and (iii) jumps back to 
Line~\ref{line:despath}. Otherwise, the UAV starts executing two concurrent 
threads (Line \ref{line:parallel}): (1) the UAV starts 
traveling to $\goal^a$, and (2) jumps back to Line~\ref{line:despath} after 
$\tau$ time units to dynamically reevaluate its current course. Concurrent to 
the above steps, each UAV periodically communicates a message to other UAVs 
every $t_c$ time units to share the set of nodes that it has visited since the 
last communication. Finally, the UAVs may periodically invoke $\NodeDivision$ 
to dynamically reorganize their goal sets. In Fig.~\ref{fig:hex}, iterative 
execution of the algorithm results in red, green, and blue paths. 

%% file: algomain.tex
\begin{algorithm}[t]
\caption{$\PathPlanning$ for UAV $a \in A$}

\begin{algorithmic}[1]


\State \label{line:nd}$V_{G}^{a} \gets \NodeDivision\left( V_{G} \right)$;

\State \label{line:despath}{\bf For each} ($v \in V_{G}^{a}$) \;
\textbf{do} \; $\hat{\vect{X}}^{a}_v \gets \; \AStar\left( v 
\right)$;

\State \label{line:PIDpath}{\bf For each} ($v \in V_{G}^{a}$) \; \textbf{do} \; 
$\vect{U}^a_v\; \gets \; \PID \left( \hat{\vect{X}}^{a}_v \right)$;

\State Compute $E_{v}^{a}$, $E_{v,\max}^a$, $E_{v,\min}^a$ from $\vect{U}^a_v$ 
using~(\ref{eq:tenergy},\ref{eq:pwe}). 

\State \label{line:goal}$\goal^{a} \gets \arg\min\Big\{E_{v}^a \, \big 
\vert \, 
{v\in V_G^a}\Big\}$;

\State \label{line:risk}$r_{\mathit{max}} \gets \max \Big\{\Risk\left( 
E_{v}^a,E_{v,max}^a,E_{v,min}^a \right) \, \big \vert \, v \in V_{G}^{a} 
-\{goal^a\}\Big\}$;

\State \label{line:tau}$\tau \gets \ReplanningInterval(r_{\mathit{max}})$;

\If {($\vect{p}_a = \vect{p}_{\goal^a}$)} \label{line:beginif}


\State $V_{G}^{a} \gets V_{G}^{a} - \left\{ \goal^a \right\}$;

\State {\bf goto}~\ref{line:despath};
\EndIf \label{line:endif}

\State \label{line:parallel}$\left\| \begin{aligned}
& \underline{\ref{line:parallel}a}: \text{Start traveling to } \goal^{a};\\
& \underline{\ref{line:parallel}b}: \text{Wait for }\tau\text{ time units, then 
{\bf goto}~\ref{line:despath};}\\
\end{aligned} \right.$
\end{algorithmic}
\label{alg:main}
\end{algorithm}

%% file: algorithm.tex
\section{Detailed Description of the Algorithm}
\label{sec:detail}

In this section, we delve into the details of Algorithm~\ref{alg:main}, 
namely, node division, path planning, energy prediction, and replanning in 
Sections~\ref{subsec:nd} -- \ref{subsec:rp}, respectively. 

\subsection{Online Node Division}
\label{subsec:nd}

In our decentralized multi-UAV path planning algorithm, each UAV initially 
chooses a group of nodes as {\em potential} goals to travel to and visit. In 
order to avoid overlaps between UAVs, we propose a distributed {\em 
synchronous}\footnote{In a synchronous distributed algorithm~\cite{l96}, each 
process sends a message to all processes and waits until it receives a message 
from all processes, including itself. When it receives all the messages, it 
continues with some local computation. Each send-receive-compute sequence is 
called a {\em round}. Synchronous algorithms typically execute multiple rounds 
before they terminate.} node division technique (see 
Algorithm~\ref{alg:NodeDivision}) that partitions the nodes in $V$.\\

\noindent {\bf Algorithm sketch. } \ The general idea of the algorithm is the 
following:

\begin{itemize}

\item First, each UAV $a \in A$ chooses a set $V_G^a$ of nodes whose traveling 
require significantly less energy than other UAVs based on its distance to the 
nodes. Then, it broadcasts $V_G^a$ to all others and receives similar messages, 
acquiring knowledge of $V_G^a$ for all $a \in A$, i.e., $\cup_{a \in A} V_G^a$.

\item Next, each UAV attempts to add the remaining nodes (i.e., nodes not in 
$\cup_{a \in A} V_G^a$) to $V_G^a$. To this end, for each such node, say $v$, 
it computes the extra energy needed to cover $V_G^a \cup \{v\}$ by computing 
the average TSP distance of these nodes.  It then broadcasts a message 
containing the extra energy needed to cover each $v$ and receives similar 
messages from other UAVs.

\item Finally, from the set of remaining nodes, each UAV chooses the nodes that 
requires the least extra energy, as compared to other UAVs.

\end{itemize}
We note that one can run $\NodeDivision$ periodically to archive better load 
balancing. We will study the impact of frequent node division in 
Section~\ref{sec:exp}.\\

\input{algond}

\noindent {\bf Detailed description.} \ The detailed steps of $\NodeDivision$ 
for each UAV $a \in A$ are listed in Algorithm~\ref{alg:NodeDivision}. Let 
$V_{G}$ denote the set of unvisited nodes. Each element of $V_G$ may be later 
added to $V_G^a$ (initially empty) for visiting. First, the UAV starts 
choosing nodes without communicating with the other UAVs (Lines 
\ref{line:ndBeginNocomLoop} -- \ref{line:bdEndNocomLoop}). In each iteration, a 
node $v_{\mathit{new}}$ that is the closest to the UAV is considered (Line 
\ref{line:ndv_new}) and we test if the traveling energy cost to 
$v_{\mathit{new}}$ by $a$ is significantly less than all other UAVs (Line 
\ref{line:ndBeginNocomcmp}). To this end, we check that the minimum traveling 
energy needed from \(\vect{p}_{a'}\), for any $a'$, to 
$\vect{p}_{v_{\mathit{new}}}$ is greater than the maximum energy needed to 
travel roundtrip from \(\vect{p}_{a}\) to \(\vect{p}_{v}\), for all $v \in 
V_{G}^{a}$. If the above condition is satisfied, we add $v_{\mathit{new}}$ to 
$V_{G}^{a}$ (Line \ref{line:ndNocomAddNode}). When all the nodes in $V_G$ are 
considered, the UAV broadcasts its goal set $V_G^a$ (Line 
\ref{line:ndBroadcastVGa}) and receives similar messages from all the UAVs. 
After receiving all messages, each UAV has full knowledge of the current set of 
goals of all other UAV and attempts to add the remaining nodes (i.e., nodes 
not in $\cup_{a \in A} V_G^a$) to $V_G^a$. To this end, we identify the  
circumcircle of the nodes in $V_G$ plus the current position of the UAV and 
compute its area $\mathcal{A}$. The nearest neighbor algorithm approximates the 
expected TSP tour length~\cite{beardwood_halton_hammersley_1959} through all 
these nodes to:
\begin{equation*}
\mathcal{L} \approx c\sqrt{n\mathcal{A}},
\end{equation*}
where $c$ is a constant and $n$ is the number of these nodes. Particularly, 
if $n=1$, then $\mathcal{A} = 0$ and if $n=2$, then the straight line between 
the two nodes is the diameter of the circumcircle. In our problem, $n = \left| 
V_{G}^{a} \right|+1$. Observe that $\mathcal{L}$ as computed in 
Lines~\ref{line:ndAn} -- 
\ref{line:ndAvgTsp} characterizes the expected total traveling energy 
cost for all the nodes in \(V_{G}^{a}\). Next, we compute the extra energy 
needed for each remaining node (Line \ref{line:ndvNotinV'1}) and include all 
these values in $S^a$, which is sent to all other UAVs (Line 
\ref{line:ndBroadcastSa}), and again, receives messages with similar information 
(Line \ref{line:ndReceiveSa}). For each unchosen node $v$ (Line 
\ref{line:ndvNotinV'2}), we check if the extra energy needed is minimum among 
that for all other UAVs (Line \ref{line:ndIfDeltaLminimum}). If so, the UAV adds 
this node to its own potential goal set $V_G^a$ (Line~\ref{line:ndComAddNode}). 
To resolve the situation that more than one UAV has minimum $\Delta L_v^a$, we 
can break the tie by choosing the UAV with minimum $\ID$ (not shown in the 
algorithm).


\subsection{Online Desirable Path Planning}
\label{OnlinePathPlanning}

For each node $v\in V_G^a$ (as computed by Algorithm~\ref{alg:NodeDivision}) on 
the $XY$-plane with the hexagon grid, we compute a {\em desirable} traveling 
path from $\vect{p}_{a}$ to $\vect{p}_{v}$, based on the well-known $\AStar$ 
algorithm~\cite{4082128}. We also consider both wind disturbance and obstacle 
bypass. We denote the set of all the hexagons by $L$ and the set of hexagons 
containing obstacles by $L_o$, where $L_o \subseteq L$. For a node $v \in V$, 
the hexagon whose center is $v$ is denoted by $l_v$. We also denote the 2D 
center position of a hexagon $l$ with $\vect{p}_{l}$. As for the UAV movement, 
each UAV moves along a sequence of centers of neighboring hexagons to reach a 
node $v$ from its current location:
$$L_{\left(a,v \right)} = l_1l_2\cdots l_n,$$
where $\vect{p}_{l_1} = \vect{p}_a$ and $l_n = l_v$. Let
$$F\left( l \right) = g\left( l \right) + h\left( l \right)$$
be the $\AStar$ {\em prediction function}, where \(g\left( l \right)\) 
represents the predicted traveling energy cost from \(\vect{p}_{a}\) to
\(\vect{p}_{l}\) through a sequence of hexagons and
\(h\left( l \right)\) is the predicted traveling energy cost from
\(\vect{p}_{l}\) to \(\vect{p}_{v}\) through a straight path.
\(F\left( l \right)\) is the total predicted traveling energy cost from
\(\vect{p}_{a}\) to \(\vect{p}_{v}\) via hexagon \(l\). The estimated wind
disturbance is modeled  as $\vect{d}(k + i) = \vect{d}(k)$, for all $i > 0$, 
where $k$ is the time index at the current position of $a$.

We compute $g(l)$ in the following way: assuming that UAV $a$ starts traveling 
from a hexagon $l'$ to another hexagon $l$ at time index $t_{l'}$, by 
Equation~\eqref{eq:tenergy}, we obtain both the time index \(t_{l}\) when \(a\) 
reaches \({l}\), such that
\begin{equation*}
t_{l}=t_{l'}+t_{( l', l)},
\end{equation*}
where $t_{( l', l)}$ is the desirable traveling time from $\vect{p}_{l'}$ to 
$\vect{p}_{l}$, and the traveling energy cost from $\vect{p}_{l'}$ to 
$\vect{p}_{l}$, i.e., $E_{( l',l)}$ under wind disturbance $\vect{d}(t_{l'})$:
\begin{equation}
\label{g(x)}
g(l) = g(l') + E^t_{( l',l)}.
\end{equation}
To compute \(h\left( l \right)\), we assume that wind disturbance is 
$\vect{d}\left(t_{l'}\right)$. By using Equation~\eqref{eq:tenergy}, traveling 
energy cost will be
\begin{equation}
h\left( l \right) = E^t_{\left( l,l_v \right)}. \label{h(x)}
\end{equation}
From \eqref{g(x)} and \eqref{h(x)}, we will get the predicted traveling energy cost $F\left(l\right)$ from $\vect{p}_a$ to $\vect{p}_v$ via hexagon $l$.

\input{algopp}

The detailed steps of our path planning algorithm are shown in 
Algorithm~\ref{alg:PathPlanning}. In Lines \ref{line:ppBeginInitgF} -- 
\ref{line:ppEndInitgF}, we first create two functions $g:L\rightarrow 
\mathbb{R}_{\geq 0}$ and $F:L \rightarrow \mathbb{R}_{\geq 0}$. We also define 
an auxiliary function $\mathit{Parent}: L \rightarrow L \cup \{\perp\}$, where 
the parent of a hexagon on a path is its predecessor. We initialize the parent 
hexagon of all the hexagons to $\perp$. We also initialize an empty {\em open 
set} $S_c$ and {\em closed set} $S_c$ for basic $\AStar$ algorithm where they 
both can contain a set of hexagons (Line~\ref{line:ppInitS_oc}). After 
initialization, we add $l_a$ into $S_o$, update $g(l_a)$ and $F(l_a)$, and set 
$t_{l_a}$ to current time index (Line \ref{line:ppAddl_a}).

Next, we enter the $\AStar$ while loop (Line \ref{line:ppBeginWhileLoop}). In 
each iteration, we first select a hexagon with minimum $F(l)$ value for 
consideration, such as $l_c$ (Line \ref{line:ppGetl_c}) and check if $l_c$ is 
the hexagon associate with node $v$ (Line~\ref{line:ppCheckl_v}). If true, our 
$\AStar$ search algorithm reaches the goal and exits the while-loop. Otherwise, 
we remove this node from $S_o$ and add it to $S_c$ 
(Lines~\ref{line:ppRemoveS_o} -- \ref{line:ppAddS_c}). Then, we start to explore 
its six neighboring hexagons one by one (denoted by $l_n$ in Line 
\ref{line:ppCheckNeighbor}). If one of them contains obstacles or is in the 
$S_c$, we skip it (Lines \ref{line:ppBeginCheckObstables} -- 
\ref{line:ppEndCheckObstacles}). Or, we add this hexagon to the $S_o$ (Line 
\ref{line:ppAddS_o}). Next, we compute the new traveling energy cost from $l_a$ 
to $l_n$, where $\mathit{Parent}(l_n) = l_c$ (Line~\ref{line:ppComputeNewMg}). 
If this new value is larger than its existed value, then we skip it, since this 
is not a better path (Line \ref{line:ppCmpNewMg}). On the contrary, we have 
found a path that cost less traveling energy to $l_n$ than its previous one. We 
set the parent hexagon of $l_n$ to $l_c$ (Line \ref{line:ppSetNewParent}), set 
$g(l_n)$ to Newgl (Line \ref{line:ppSetNewMg}), compute its new $F(l_n)$ (Line 
\ref{line:ppSetnewMf}) and the predicted time index when the UAV reaches $l_n$ 
(Line \ref{line:ppSetNewMt}). Once $S_o = \varnothing$, we leave the while-loop 
and generate waypoints from $l_a$ to $l_v$ (Line \ref{line:ppBuildPath}).

Finally, we generate the desirable states of the UAV at each time index from 
$l_a$ to $l_v$ by using the polynomial motion planning 
method~\cite{babel2013three,caubet2015motion,caubet2015inverse}. At each time 
index, the desirable 
position, velocity and acceleration of a UAV are generated by polynomial 
fitting functions. The desirable Eular angles and angular velocities are all 
$\vect{0}$. The result is shown as:
\begin{equation*}
\hat{\vect{X}}_v^{a} = \hat{\vect{x}}_a\left(t_{l_a}\right) 
\hat{\vect{x}}_a\left(t_{l_a}+1\right) \hat{\vect{x}}_a\left(t_{l_a}+2\right) 
\cdots \hat{\vect{x}}_a\left(t_{l_v}\right).
\end{equation*}
The above equation is has the same meaning as \eqref{Xhat_v^a}, since
$t_{l_a}=t_{\vect{p}_a}$ and $t_{l_v}=t_{\vect{p}_a} + t_{\left( \vect{p}_{a}, 
\vect{p}_{v} \right)}$.

\subsection{Online PID-based Energy Prediction}
\label{OnlinePID-basedEnergyPrediction}

In the previous subsection, for each UAV $a\in A$, we obtained a sequence of 
desirable states through all the waypoints from $\vect{p}_a$ to $\vect{p}_v$, 
for each $v \in V_G^a$. Now, we are going to generate the predicted input 
$\vect{u}$ from each pair of neighboring desirable states in 
$\hat{\vect{X}}_v^{a}$ based on the wind disturbance $\vect{d}(t_{\vect{p}_a})$ 
at current time index. Then, we can compute the predicted traveling energy cost 
with all the inputs. Since the dynamics of quadrotors is nonlinear, and the 
position control (i.e. thrust force) and attitude control (i.e. moments in 
three dimensions) are coupled with each other, it is difficult to use a PID 
controller directly. However, one can decouple these two control designs using 
the technique in~\cite{zuo2010trajectory} and linearize each controller design. 
For example, assuming that $\vect{p}_c$ is the desirable position and 
$\vect{p}$ is the current position, in order to design the position control, we 
can build the position closed-loop equation as follows:
\begin{equation}
\ddot{\vect{p}}_e+K_{d}\dot{\vect{p}}_e+K_{p}\vect{p}_e=0 \label{PositionClosedLoopEquation}
\end{equation}
where $\vect{p}_e=\vect{p}_c-\vect{p}$; $\dot{\vect{p}}_e$ and 
$\ddot{\vect{p}}_e$ are first and second order derivatives of $\vect{p}_e$, 
respectively, and $K_{d}$ and $K_{p}$ are both $3\times3$ positive definite 
matrices. Since $\ddot{\vect{p}}$ is related to the UAV thrust force 
directly, Equation~\eqref{PositionClosedLoopEquation} can be expanded as:
\begin{equation}
\ddot{\vect{p}}=\ddot{\vect{p}}_c+K_{d}\left( \dot{\vect{p}}_c - \dot{\vect{p}} \right)+K_{p}\left( \vect{p}_c - \vect{p} \right).
\end{equation}
The attitude control design is similar to position control design as above.

We utilize the results in~\cite{zuo2010trajectory} as follows. Let 
$\vect{x}_a(t_{\vect{p}_a})$ be the current state which is the same as 
$\hat{\vect{x}}_a\left(t_{\vect{p}_a}\right)$. We generate 
$\vect{u}_a(t_{\vect{p}_a})$ from $\vect{x}_a(t_{\vect{p}_a})$ and 
$\hat{\vect{x}}_a(t_{\vect{p}_a}+1)$ by using the decoupled PID control 
prediction. Besides, we obtain predicted states $\vect{x}_a(t_{\vect{p}_a}+1)$ 
from $\vect{x}_a(t_{\vect{p}_a})$ and $\vect{u}_a(t_{\vect{p}_a})$. Typically, 
there is some difference between $\vect{x}_a(t_{\vect{p}_a}+1)$ and 
$\hat{\vect{x}}_a(t_{\vect{p}_a}+1)$. We continue this step from desirable 
states $\hat{\vect{x}}_a\left(t_{\vect{p}_a}\right)$ to 
$\hat{\vect{x}}_a\left(t_{\vect{p}_a} + t_{\left( \vect{p}_{a}, \vect{p}_{v} 
\right)} \right)$. Finally, we obtain a sequence of predicted inputs as follows:
\begin{eqnarray*}
\vect{U}_v^{a} &=& \vect{u}_a\left(t_{\vect{p}_a}\right) \vect{u}_a\left(t_{\vect{p}_a}+1\right) \vect{u}_a\left(t_{\vect{p}_a}+2\right)\\
&&\cdots \vect{u}_a\left(t_{\vect{p}_a} + t_{\left( \vect{p}_{a}, \vect{p}_{v} \right)}-1\right).
\end{eqnarray*}

Next, we compute the predicted traveling energy cost from the above input 
sequence. For each predicted input vector $\vect{u}_a \in \vect{U}_v^a$, we 
first convert it into rotational speed of the four propellers by 
\eqref{InputComputation}, then we obtain the piece-wise traveling energy 
cost by Equation \eqref{eq:tenergy} with $t_{\left( \vect{p}_{i},\vect{p}_{j} 
\right)}$ being the sampling time $t_s$. Next, we summarize all the piece-wise 
traveling energy cost together to get the predicted traveling energy cost 
\(E_{v}^{a}\) from $\vect{p}_a$ to $\vect{p}_v$ by UAV $a$. Each UAV computes 
\(E_{v}^{a}\) for all the nodes in $V_{G}^{a}$. Besides computing 
\(E_{v}^{a}\) for each node $v$ in \(V_{G}^{a}\), UAV \(a\) also computes the 
maximum and minimum traveling energy cost with the same path, such that 
$E_{v,\max}^a$ and $E_{v,\min}^a$ with the wind disturbance $\vect{d}(k)$ being 
$\vect{d}_{\max}(k)$ and $\vect{d}_{\min}(k)$, respectively. Finally, the UAV  
chooses the node \(goal^{a}\), such that \(goal^{a} = \arg{\min_{v\in 
V_G^a}\{E_{v}^a\}}\), as the goal to travel to and service, as described in 
Algorithm~\ref{alg:main}.

\subsection{Online Replanning}
\label{subsec:rp}

In order to achieve more accurate path planning and energy prediction in the 
presence of time variant and un-predictable disturbances,  we apply periodic 
replanning. To this end, for each node in $V_G^a$ except $goal^a$, we use a risk 
number $r$ to represent the risk that this node could become the one that has 
minimum \(E_{v}^{a}\) \cite{bmnlkp16}. The risk number is defined as follows:
\begin{equation}
r_{v} = \frac{\max\left\{ E_{goal^a,\max}- E_{v,\min},0 \right\}}{E_{v,\max} - E_{goal^a,\min}}\ \ \ \forall v \in V_{G}^{a}.v \neq goal^a.
\end{equation}
We choose the maximum $r_v$ which is denoted by $r_{max}$ to compute the 
replanning interval defined as:
\begin{equation}
\tau = \alpha\left( 1 + \frac{\beta}{r_{max}} \right),
\end{equation}
where \(\alpha\) is the minimum replanning interval and \(\beta\) is a proper 
positive nature number. Then, the next planning time will be
\begin{equation}
t_{a,k + 1}^{\text{plan}} = t_{a,k}^{\text{plan}} + \tau.
\end{equation}

%% file: algond.tex
\begin{algorithm}[t]
\caption{$\NodeDivision$ for UAV $a \in A$}

\hspace*{\algorithmicindent} \textbf{Input:} $V_G$

\begin{algorithmic}[1]

\State $V_G^a \gets \varnothing$; \label{line:ndinit}
\BeginForeach{$v \in V_{G}$} \label{line:ndBeginNocomLoop}

\State $v_{\mathit{new}} \gets \arg\min 
\Big\{\left\|\vect{p}_a-\vect{p}_v\right\| \; \Big\vert \; v\in 
V_G-V_G^a\Big\}$; \label{line:ndv_new}

\If{$(\forall a'\in A -\{a\}: 2\sum_{v \in V_{G}^{a}\cup\left\lbrace 
v_{\mathit{new}}\right\rbrace}E_{v,\max}^{a} < E_{v_{\mathit{new}}, 
\min}^{a'})$}\label{line:ndBeginNocomcmp}
$~~V_G^a\gets V_G^a\cup\left\lbrace v_{\mathit{new}} \right\rbrace$; 
\label{line:ndNocomAddNode}

\EndIf \label{line:ndEndNocomcmp}
\EndForeach  \label{line:bdEndNocomLoop}

\State \textbf{Broadcast} $\big(V_G^a\big)$;\label{line:ndBroadcastVGa}

\State $V' \gets$ \textbf{Receive} $\left(\cup_{a' \in A} \{V_G^{a'}\}\right)$;\label{line:ndReceiveVGa}

\State Let $\mathcal{A}$ be the area of the circumcircle of $V_G^a \cup 
\{\vect{p}_a\}$; \label{line:ndAn}

\State $\mathcal{L} \gets c\sqrt{\left(\left| V_G^a 
\right|+1\right)\mathcal{A}}$;\label{line:ndAvgTsp}

\State $S^a\gets \varnothing$;\label{line:ndInitSa}

\BeginForeach{$v \in V_G - V'$}\label{line:ndvNotinV'1}


\State Compute $\mathcal{L}'$ for $V_G^a\cup \left\lbrace v \right\rbrace$ as 
in Lines 
\ref{line:ndAn} -- \ref{line:ndAvgTsp};

\State $\Delta \mathcal{L}_v^a\gets \mathcal{L}'-\mathcal{L}$; 
\label{line:ndDeltaL}

\State $S^a \gets S^a \cup \left\lbrace \Delta \mathcal{L}_v^a 
\right\rbrace$;\label{line:ndAddDeltaLToSa}


\EndForeach

\State \textbf{Broadcast} $\left(S^a\right)$;\label{line:ndBroadcastSa}

\State \textbf{Receive} $\left(\cup_{a' \in A} \{\Delta \mathcal{L}_v^{a'} 
\}_{v \in V_G - V'}\right)$;\label{line:ndReceiveSa}

%
%

\BeginForeach{$v\in V_G - V'$}\label{line:ndvNotinV'2}

\If{$\Delta \mathcal{L}_v^a = \min\left\lbrace \Delta \mathcal{L}_v^{a'} \, 
\big\vert \, a' \in 
A\right\rbrace $}\label{line:ndIfDeltaLminimum}
	\State $V_G^a\gets V_G^a\cup\left\lbrace v \right\rbrace$;\label{line:ndComAddNode} 	
\EndIf
\EndForeach

\end{algorithmic}
\label{alg:NodeDivision}
\end{algorithm}

%% file: algopp.tex
\begin{algorithm}[t]
\caption{$\AStar$ function for UAV $a \in A$ and node $v \in V_G^a$ }


\begin{algorithmic}[1]
\BeginForeach{$l\in L$}\label{line:ppBeginInitgF}
\State $g\left(l\right), F\left(l\right)\gets 
+\infty$; $Parent\left(l\right)\gets \perp$;
\label{line:ppInitgF}
\EndForeach\label{line:ppEndInitgF}

\State $S_o, S_c \gets \varnothing$;\label{line:ppInitS_oc}

\State $g\left(l_a\right)\gets 0$; $F\left(l_a\right)\gets 
g\left(l_a\right)+h\left(l_a\right)$; $t_{l_a}\gets k$; $S_o\gets \{l_a\}$; 
\label{line:ppAddl_a}

\While{$(S_o\ne \varnothing)$} \label{line:ppBeginWhileLoop}

\State $l_c\gets \arg\min\left\lbrace F\left(l\right) \, \big\vert \, l\in S_o 
\right\rbrace$; \label{line:ppGetl_c}

\If{$(l_c =l_v)$} \label{line:ppCheckl_v}
\State \textbf{break};
\EndIf

\State $S_o\gets S_o - \left\lbrace l_c \right\rbrace$; \label{line:ppRemoveS_o}

\State $S_c\gets S_c \cup \left\lbrace l_c \right\rbrace$; \label{line:ppAddS_c}

\BeginForeach{$l_n \in \mathit{Neighbor}(l_c)$} \label{line:ppCheckNeighbor}
\If{$(l_n \in L_o \cup S_c)$}\label{line:ppBeginCheckObstables}
		\State \textbf{continue};
\EndIf\label{line:ppEndCheckObstacles}

\State $S_o \gets S_o \cup \left\lbrace l_n \right\rbrace$;\label{line:ppAddS_o}

\State Newgl $\gets g\left(l_c\right)+E_{\left( 
l_c,l_n \right)}$; \label{line:ppComputeNewMg}

\If{(Newgl $\ge g\left(l_n\right))$} \label{line:ppCmpNewMg}
\State \textbf{continue};
\EndIf 

\State $\mathit{Parent}\left(l_n\right)\gets l_c$; 
\label{line:ppSetNewParent}

\State $g\left(l_n\right)\gets$ Newgl; \label{line:ppSetNewMg}

\State $F\left(l_n\right)\gets g\left(l_n\right)+h\left(l_n \right)$;
\label{line:ppSetnewMf}

\State $t_{l_n}\gets t_{l_c}+t_{\left( l_c,l_n \right)}$;\label{line:ppSetNewMt}
\EndForeach
\EndWhile

\State $\mathit{WayPoints}_{(a,v)} \gets \mathsf{BuildPath}(l_v, 
\mathit{Parent})$; \label{line:ppBuildPath}

\State $\hat{\vect{X}}_v^{a} \gets \mathsf{GenerateDesirableStates}( 
\mathit{WayPoints}_{(a,v)})$; \label{line:ppGenerateDesirableStates}
\end{algorithmic}
\label{alg:PathPlanning}
\end{algorithm}

%% file: experiments.tex
\section{Evaluation}
\label{sec:exp}

In order to analyze the performance of Algorithm~\ref{alg:main}, we have 
conducted a rigorous set of simulations.

\begin{figure}[t]
 \centering
 \includegraphics[scale=.75]{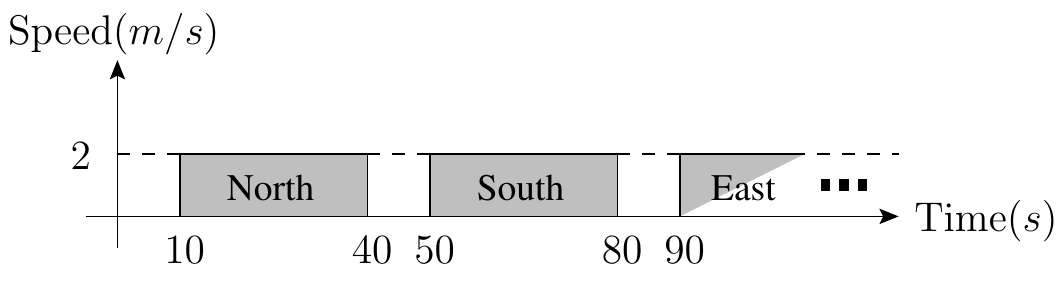}
 \caption{Low wind pattern.}
 \label{fig:wind}
\end{figure}

\subsection{Experimental Settings}

We have implemented all the algorithms in MATLAB. In our simulations, we 
consider the following {\em parameters}:

\begin{itemize}
\item The size of the flight area ranges over $52 \times 30$, $78 \times 45$, 
and $104 \times 60$ squared meters. We fix the size of hexagons with side 
lengths of $1m$.

\item The number of nodes in the area varies between 10 -- 50.

\item The number of UAVs ranges over 1 -- 8.

\item The wind disturbance is $2m/s$ or $8m/s$ to represent {\em low} and 
{\em high} winds, respectively. For the duration of experiments, wind blows for 
$30s$ within $40s$ time intervals randomly in one of the north, south, east, and 
west directions. There is $10s$ no wind between any two consecutive wind events 
(see Fig.~\ref{fig:wind}). Note that given the direction of wind, angle 
$\theta$ in Equation~\eqref{eq:tenergy} can be computed from the direction of a 
UAV.

\item The arm length of all UAVs are $8.6cm$ and their mass is $0.18kg$.

\item The PID sampling period is fixed at $0.005s$.

\item We distinguish four cases based on enabling and disabling node redivision 
(denoted $\ND$) and online replanning (denoted $\OP$). The four cases are 
denoted by $[\ND, \OP]$, $[\ND, \neg \OP]$, $[\neg \ND, \OP]$, and $[\neg \ND, 
\neg\OP]$. In cases with $\neg \ND$, all the UAVs do a one-time node division 
in the beginning (i.e., $k=0$). And, for the cases with $\ND$, node division is 
applied with different time periods $10$ -- $80$ seconds.

\end{itemize}

Our sole {\em metric} for measurement is the {\em actual} global traveling 
energy consumption in Joules. To gain statistical confidence, for each 
experiment, we generate 100 different random data sets, where the positions of 
the nodes and the UAVs are independent and identically scattered in the flight 
area. This will ensure $95\%$ confidence interval. We calculate the average 
energy consumption of all the experiments.
 
\subsection{Analysis of Results}

\subsubsection{Comparing Algorithm~\ref{alg:main} with a greedy algorithm}

We first focus on comparing the performance of our algorithm with a best-effort 
algorithm, where each idle UAV chooses to visit the closest unvisited node. We 
fix the number of UAVs to 4 and wind to $2m/s$.

\begin{figure}[t]
\centering

\begin{subfigure}[t]{0.49\columnwidth}
\centering
\includegraphics[scale=0.16]{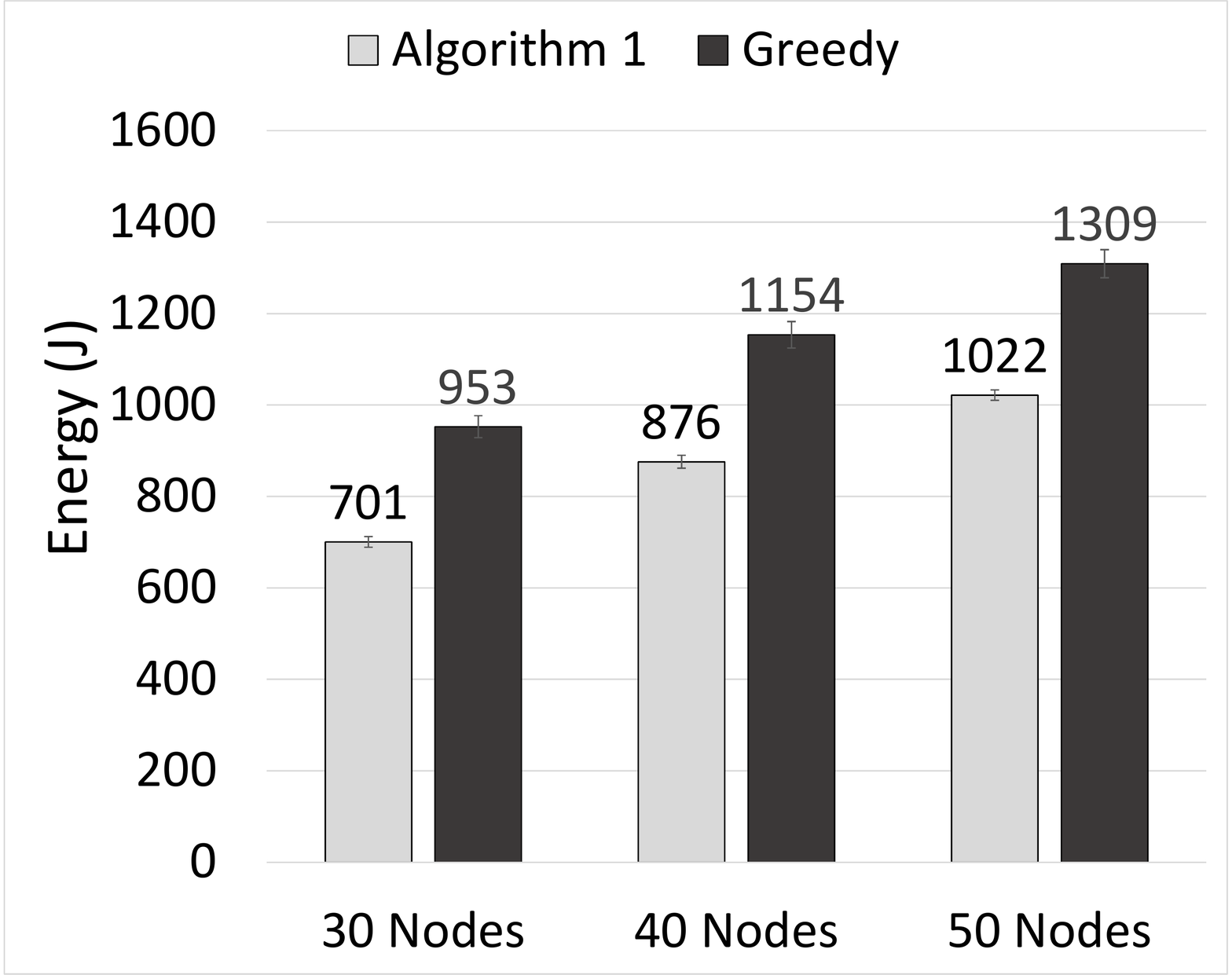}
\caption{Fixed area $52 \times 30m^2$.}
\label{Fig_Greedy_Algorithm1_Nodes}
\end{subfigure}
\hfill
\begin{subfigure}[t]{0.49\columnwidth}
\centering
\includegraphics[scale=0.16]{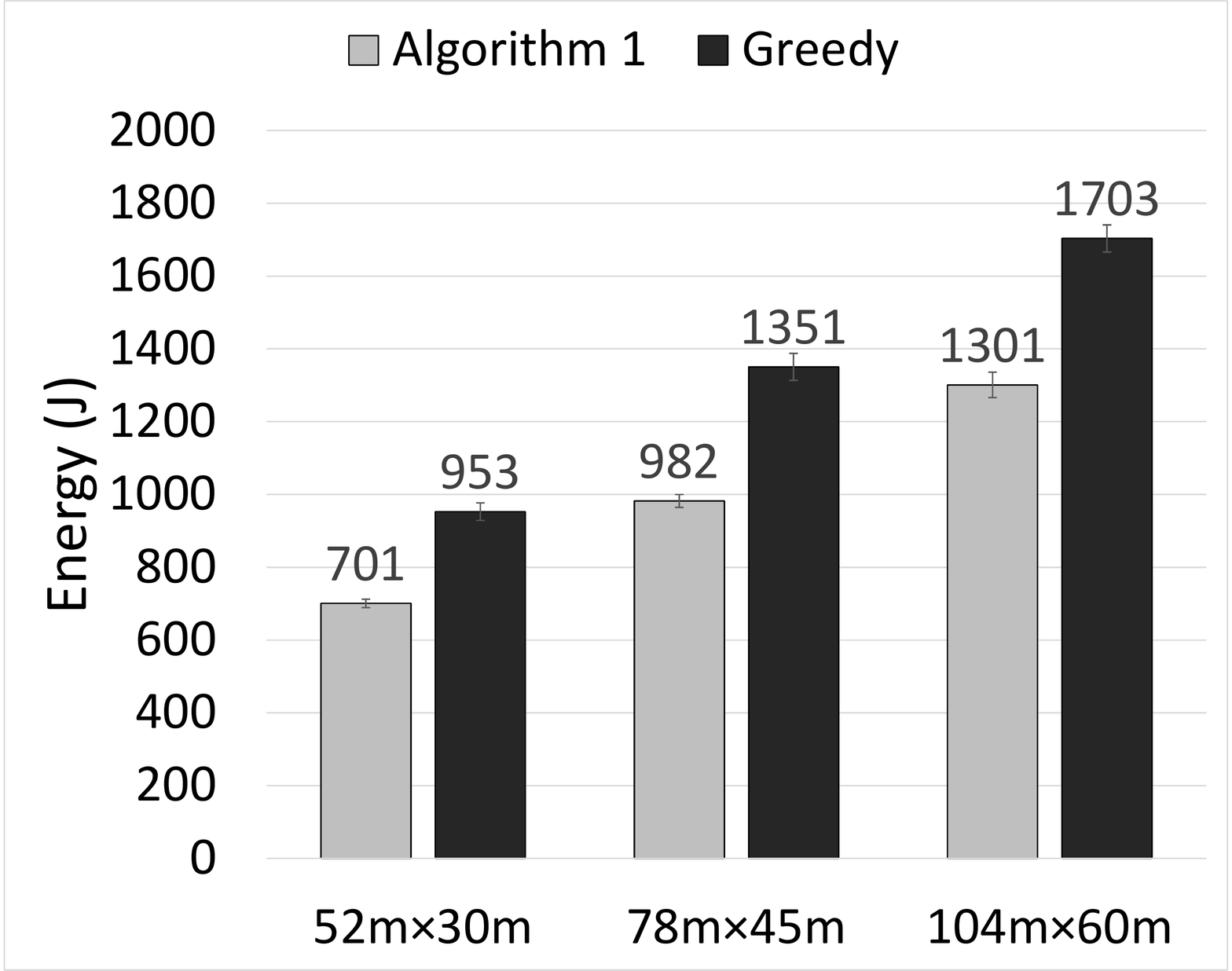}
\caption{Fixed number of nodes 30.}
\label{Fig_Greedy_Algorithm1_Area}
\end{subfigure}

\caption{Comparing Algorithm~\ref{alg:main} with the greedy algorithm for 4 
UAVs and node redivision period $10s$.}
\label{fig:gvsalg1}
\end{figure}

\paragraph{Fixed area/variable number of nodes} Here, we compare two strategies: 
the greedy algorithm and Algorithm~\ref{alg:main} with node redivision interval 
of $10s$, i.e., function $\NodeDivision$ is invoked every $10s$. The flight 
area is fixed at $52 \times 30m^2$. Figure~\ref{Fig_Greedy_Algorithm1_Nodes} 
shows the performance for 30, 40, and 50 nodes. As can be seen, 
Algorithm~\ref{alg:main} clearly outperforms the greedy algorithm by almost 
$24\%$ energy savings on average. The main reason is that in our approach a UAV 
can keep a set of nodes in its close vicinity and other UAVs that are in 
relatively longer distances will not travel to one of those nodes. Observe that 
the performance drops a little bit as the number nodes grow. This is because 
the area becomes more dense and the assignment of nodes to UAVs in both 
algorithms become less crucial. 

\begin{figure*}[h]
\centering

\begin{subfigure}[t]{0.49\columnwidth}
\centering
\includegraphics[scale=0.17]{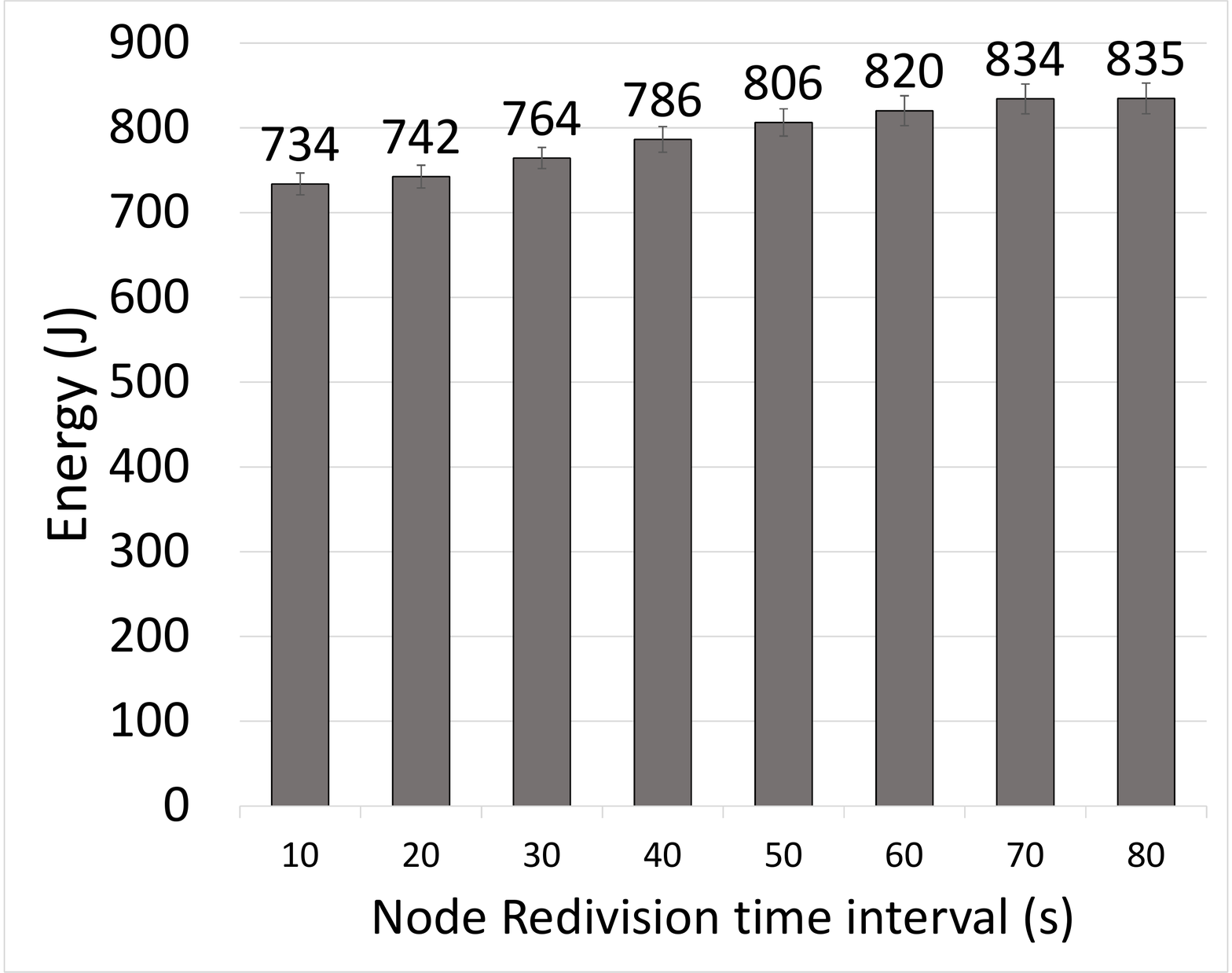}
\caption{Variable time period.}
\label{Fig_Frequency_CI}
\end{subfigure}
\hfill
\begin{subfigure}[t]{0.49\columnwidth}
\centering
\includegraphics[scale=0.17]{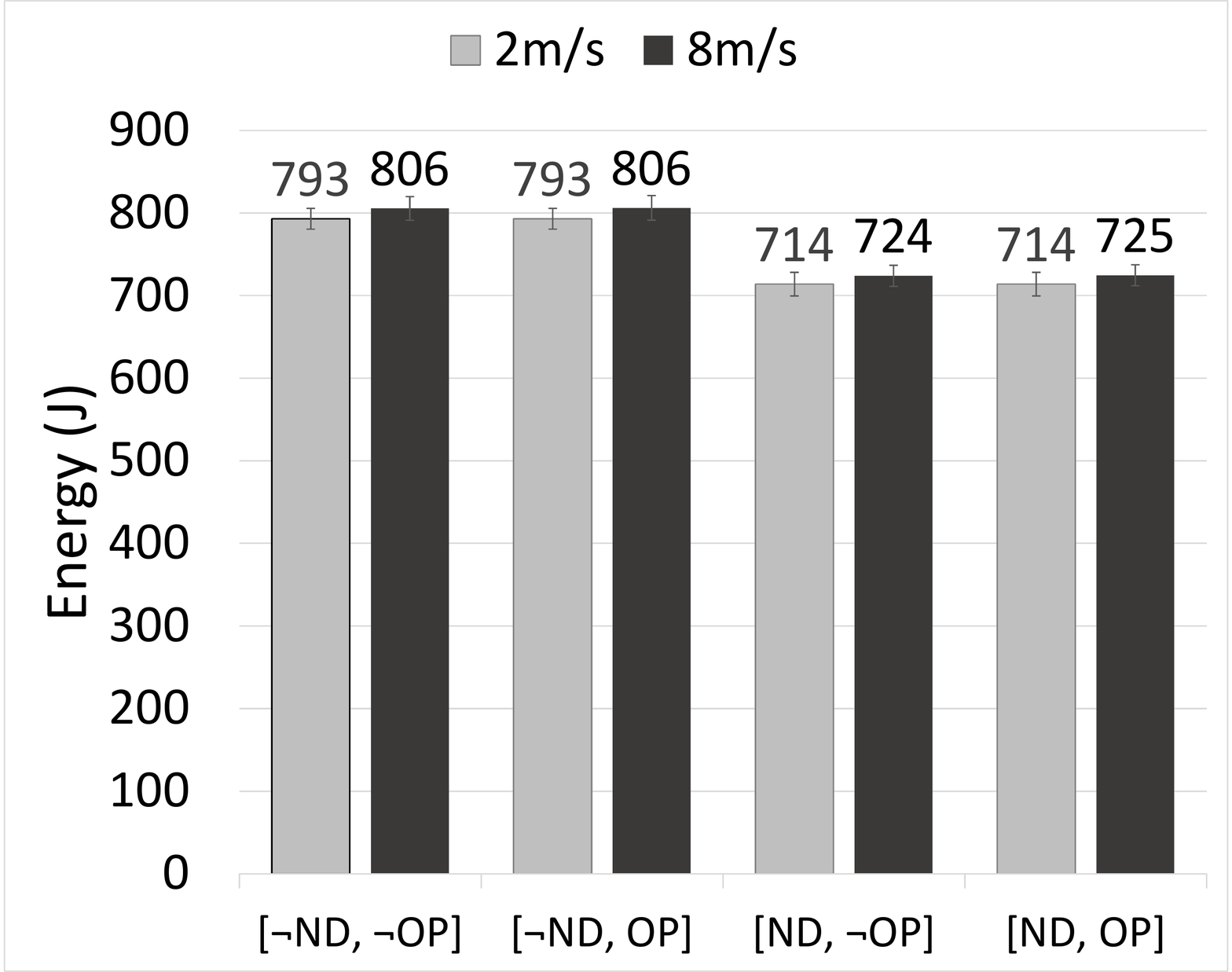}
\caption{Wind disturbance.}
\label{Fig_windspeed}
\end{subfigure}
\hfill
\begin{subfigure}[t]{0.49\columnwidth}
\centering
\includegraphics[scale=0.17]{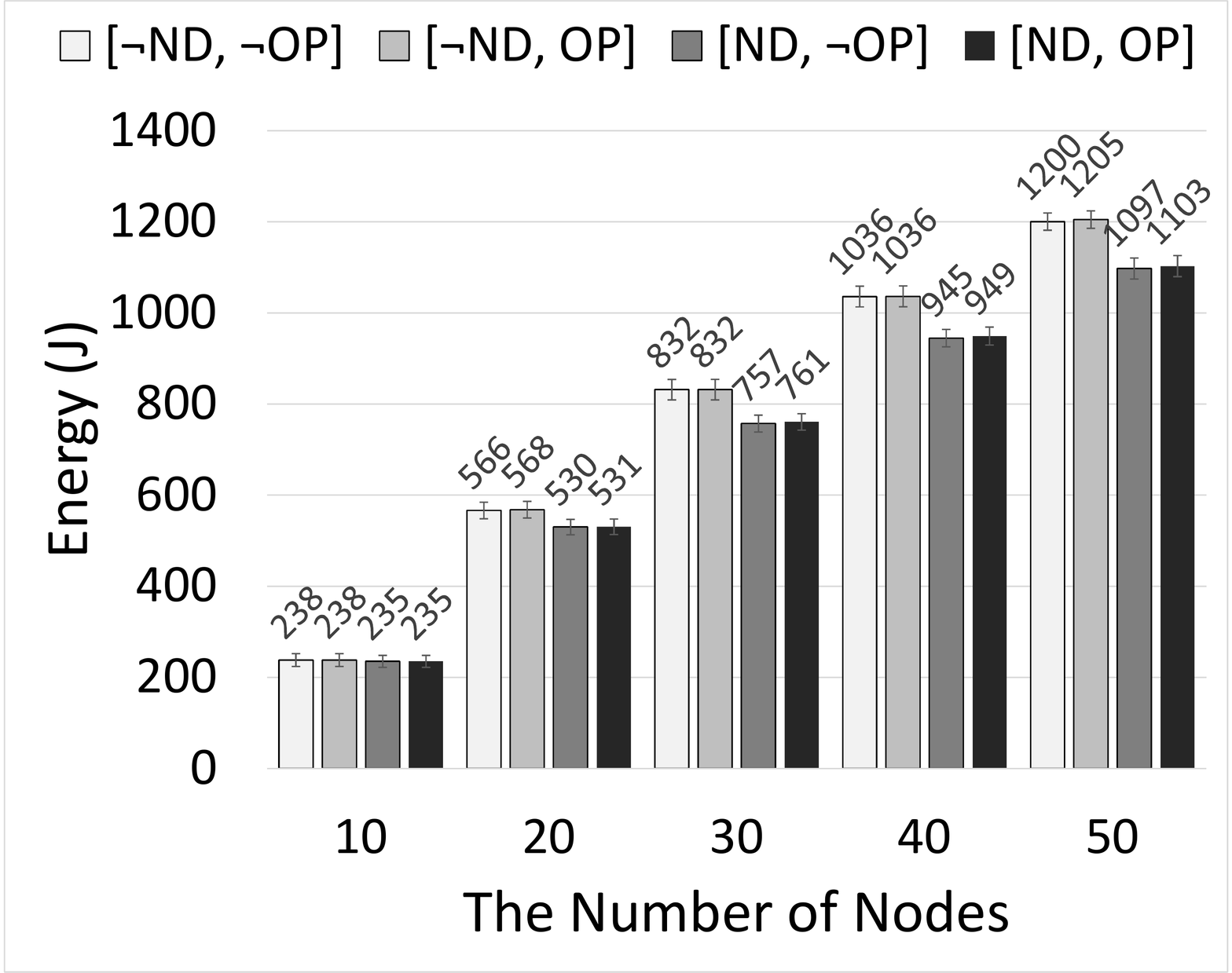}
\caption{The number of nodes.}
\label{Fig_Node}
\end{subfigure}
\hfill
\begin{subfigure}[t]{0.49\columnwidth}
\centering
\includegraphics[scale=0.17]{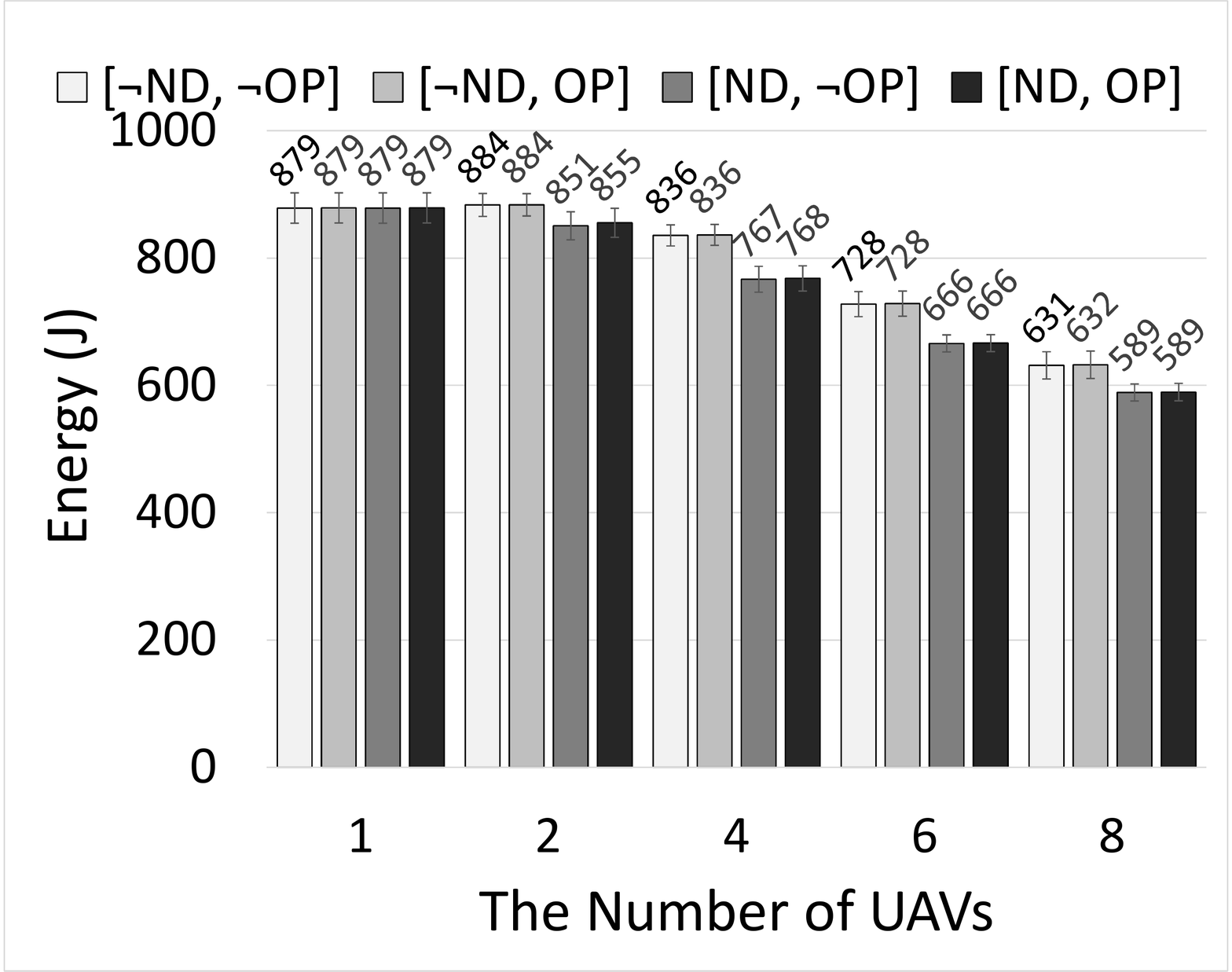}
\caption{The number of the UAVs.}
\label{Fig_UAV}
\end{subfigure}

\caption{The impact of node redivision.}
\label{fig:ImpactND}
\end{figure*}

\paragraph{Fixed number of nodes/variable area} Here, we compare the greedy 
algorithm with Algorithm~\ref{alg:main} by fixing the number of nodes to 30 and 
changing the size of flight area. The rest of the parameters are identical to 
those of the previous section. We consider area sizes $52m \times 30m$, $78m 
\times 45m$, and $104m \times 60m$. The performance is shown in 
Figure~\ref{Fig_Greedy_Algorithm1_Area}. Our Algorithm~\ref{alg:main} always has 
about $26\%$ energy saving when compared with the greedy algorithm.

\subsubsection{The Impact of Node Redivision}

We now focus on the impact of function $\NodeDivision$ and the interval of its 
invocation on the performance of Algorithm~\ref{alg:main}. Here, all 
experiments are in $52 \times 30m^2 $ area. 

\paragraph{Node redivision period} Now, we test the impact of the time interval 
of node redivision. We fix the number of nodes to 30, the number of UAVs to 4, 
the disturbance to the low wind, and increase the node redivision period from 
$10s$ to $80s$. In Fig.~\ref{Fig_Frequency_CI}, the average traveling energy 
cost grows as the the node redivision interval increases. The reason is that 
function $\NodeDivision$ utilizes TSP-length estimation which may be less 
accurate than the actual UAV path. Thus, adjusting the estimation periodically 
is beneficial. In addition, if we continue increasing the node redivision 
interval, the average traveling energy cost will approach to that with only 
one-time node division at the beginning. It can be seen that $10s$ period is 
$13\%$ more efficient than $80s$ interval.

\paragraph{Wind speed} Now, we change the wind disturbance by fixing 30 nodes 
with four UAVs. We also distinguish four cases based on enabling $\ND$ and 
$\OP$. Figure~\ref{Fig_windspeed} shows the simulation results. As can be seen, 
high wind will increase the overall traveling energy cost. Enabling $\OP$ does 
not show performance benefits as compared to node redivision in low wind. In 
high wind, the effect is hard to predict. Sometimes, it can reduce the overall 
traveling energy cost, but sometimes it makes the results even worse. Online 
replanning tends to choose the node with minimum predicted traveling energy as 
the goal to travel to, but the local optimum choice does not always lead to 
global optimality. Also, although online replanning may change the goal node 
once in a while, its impact in cases with large number of nodes and UAVs is 
almost negligible. This is, however, not the case for smaller number of 
nodes and UAVs. We will discuss this in Section~\ref{subsec:op}. On the 
contrary, node redivision always give better average results.

\paragraph{Number of nodes in flight area} Figure~\ref{Fig_Node} shows our 
results for 10 -- 50 nodes. We fix four UAVs and low wind. As can be seen, the 
performance of node redivision improves as the number of nodes grow. On the 
contrary, with a small number of nodes, node redivision usually gives similar 
results to one-time node division, as the nodes that are assigned to each UAV 
is small. Moreover, since we apply random distribution of nodes in the area, the 
average distance between nodes is large when the number of nodes is small. When 
a UAV moves closer to its goal, its distance to other nodes may grow and other 
UAVs have less chance to choose this node.

\paragraph{The number of UAVs} Figure~\ref{Fig_UAV} shows our results for 1 -- 
8 UAVs while fixing 30 nodes and low wind. As can be seen, when the number of 
UAVs is small, node redivision has little effect, which is expected. As the 
number of UAVs grow, the performance of the node redivision improves. However, 
when there are too many UAVs, the effect of node redivision again fades 
away, which is also expected. Thus, there is a sweet spot in choosing the 
number of UAVs proportional to the area size and the number of nodes in the 
area.

\begin{figure}[t]
\centering
\begin{subfigure}[t]{0.49\columnwidth}
	\centering
	\includegraphics[scale=0.49]{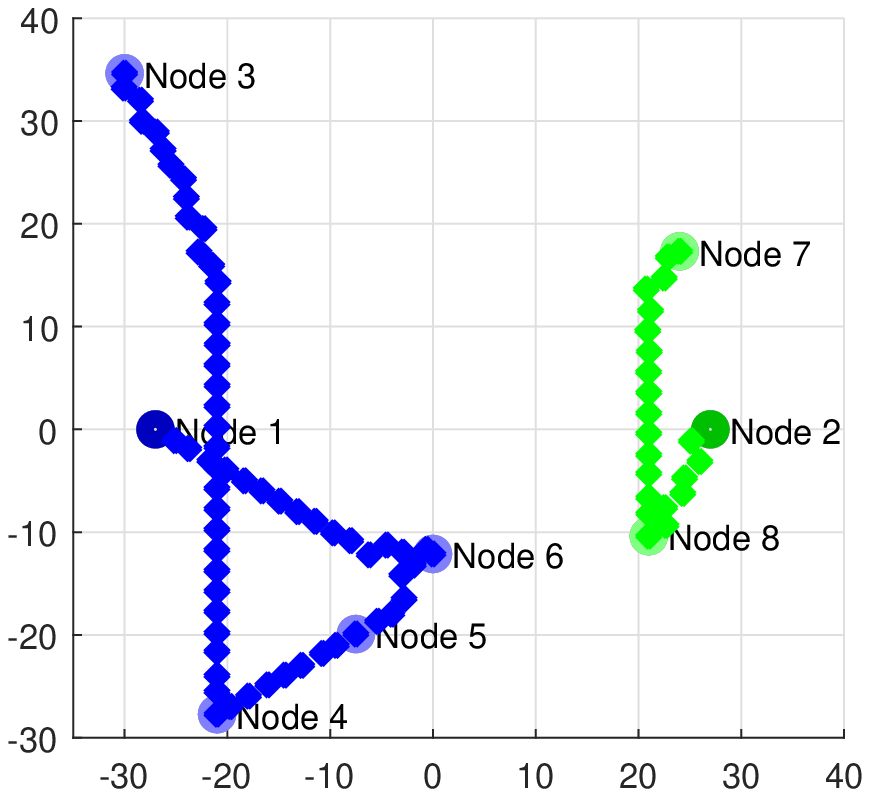}
	\caption{Disabled online replanning.}
	\label{Fig_Disable_Replannning}
	\end{subfigure}
\hfill
\begin{subfigure}[t]{0.49\columnwidth}
	\centering
	\includegraphics[scale=0.49]{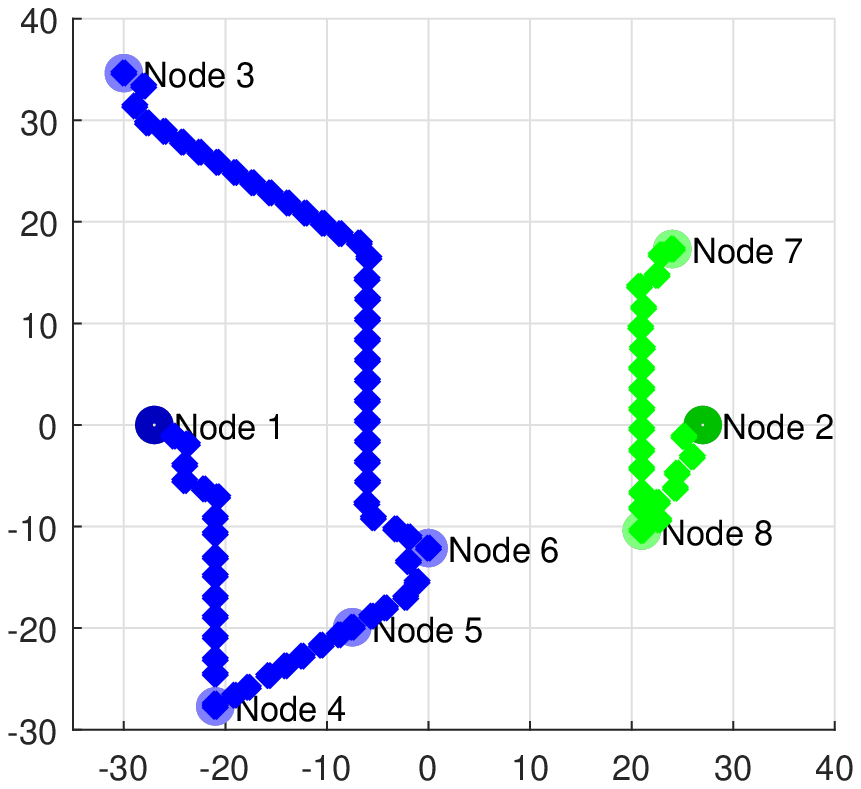}
	\caption{Enabled online replanning.}
	\label{Fig_Enable_Re-plannning}
\end{subfigure}

\caption{The impact of online replanning.}
\label{fig:EffectOP}
\end{figure}

\subsubsection{The Impact of Online Replanning}
\label{subsec:op}

There exist cases, where online replanning results in better performance. 
Figures~\ref{Fig_Disable_Replannning} and~\ref{Fig_Enable_Re-plannning} show 
the flight paths of two UAVs located at Node 1 and Node 2. There is only 
a one-time initial node division. As we can be seen, the overall traveling 
energy cost with disabling and enabling online replanning is about $378J$ 
and $372J$, respectively. When UAV 1 leaves its depot, there is a $8m/s$ 
wind from north to south. If we enable online replanning, the UAV changes its 
goal to Node 4 instead of Node 6 and the overall traveling energy cost will 
decrease. This result is consistent with the technique in~\cite{bmnlkp16}, 
where online replanning is applied for single UAV scenarios.

%% file: related.tex
\section{Related Work}
\label{sec:related}



The {\em vehicle routing problem} (VRP)~\cite{tv02} is generally concerned with 
the optimal design of routes by a fleet of vehicles to service a set of 
customers by minimizing the overall cost, usually the travel distance for the 
whole set of routes. To our knowledge, the body of work on distributed VRP is 
limited to the work in~\cite{pfb11}, where the authors propose partitioning 
policies for adaptive vehicle routing in a dynamic environment, where travel 
and service costs can change. In distributed vehicle routing 
approximation~\cite{kmb17}, the authors propose a distributed approximation 
algorithm to solve VRP. The DRONA framework~\cite{dsyqs17} provides 
provably correct distributed path planning for mobile robots. Heuristic 
algorithms has been widely used to solve VRP too. For example, particle swarm 
optimization (PSO)~\cite{mirhassani2011particle} generates new search steps by 
learning from current and historical best solutions with fast convergence. 
Ant colony optimization (ACO)~\cite{li2009ant} employs a pheromone model, 
which builds an initial feasible solution for VRP and then a tabo search 
is used to improve the current solution to local optimal. In these approaches, 
there is no notion of physical disturbances.


UAV flight and control have been widely explored. In~\cite{zuo2010trajectory}, 
 the authors survey the dynamics of quadrotors under physical constraints. 
The work in~\cite{bmnlkp16} studies disturbance-aware online path planning. In 
their model, a single UAV needs to visit multiple nodes. They employ 
model-predictive control (MPC) to predict the traveling energy cost to each node 
under current wind disturbance and choose the node with minimum predicted 
traveling energy cost to travel. Since wind is time variant, the algorithm runs 
MPC prediction at a relatively high rate to adjust the node choice. The paper 
uses an online self-triggered scheduling technique to dynamically reschedule 
the next replanning time based on the latest MPC prediction result.

%% file: concl.tex
\section{Conclusion and Future Work}
\label{sec:concl}

We introduced a decentralized and online path planning technique for a network 
of quadrotor UAVs in the presence of weather disturbances, where the vehicles 
are expected to collaboratively visit a set of nodes scattered in a 
2D area. Each UAV will have to spend energy to reach these nodes and defy 
environment disturbances. Our approach consists of four main components (i) a 
distributed algorithm that periodically divides the remaining unvisited nodes 
among the UAVs, (ii) a local (i.e., UAV-level) $\AStar$-based algorithm that 
computes the {\em desirable} path for each UAV to reach the nodes assigned to 
it, (iii) a local PID controller that predicts the UAV inputs, and (iv) a 
planner that computes the predicted energy and replanning time period. We showed 
that our techniques results in significant energy savings as compared to an 
intuitive best-effort algorithm.

We are currently implementing our technique in a real network of UAVs on the px4 
flight stack. There are many interesting research avenues to explore on top of 
our results. A natural extension is to consider a 3D model, where obstacles 
can be circumvented by flying around or over them. These choices have different 
energy consumption profiles. Another problem is to consider environments with 
dynamic obstacles, vehicles, and goals. One may also take other environmental 
disturbances such as the heat into account.

\section{Acknowledgement}

We would like to thank Huawei Zhu for sharing her insights on PID controller 
decoupling techniques. 